\documentclass{emulateapj} 
\usepackage[dvips]{color,epsfig} 
\usepackage{apjfonts}

\begin{document}

\title{Low-Redshift Damped Lyman Alpha Galaxies
Toward  the Quasars B2 0827+243, PKS 0952+179, PKS 1127$-$145, and  
PKS 1629+120\altaffilmark{1}}  \altaffiltext{1}{Based on observations
obtained with the 3.5m WIYN Telescope on Kitt Peak, operated
for the NSF by the Association of Universities for Research in 
Astronomy (AURA), Inc. (WIYN is a joint facility of University of Wisconsin,
Indiana University, Yale University,  and NOAO), the 
Infrared Telescope Facility, which is operated by the University of 
Hawaii under a cooperative agreement with the National Aeronautics
and Space Administration, the Hiltner
2.4m Telescope on Kitt Peak, operated by MDM  Observatory (this is a
joint facility of University of Michigan, Dartmouth College, Ohio
State University, and Columbia University), and the 3.6m European
Southern Observatory New Technology Telescope on La Silla, Chile.  }

\author{Sandhya M. Rao\altaffilmark{2}, Daniel B. Nestor, David
A. Turnshek,} 

\affil{Department of Physics \& Astronomy, University of Pittsburgh,
Pittsburgh, PA 15260}

\author{Wendy M. Lane,}

\affil{US Naval Research Lab, 4555 Overlook
Ave, SW, Code 7600A, Washington, DC 20375}

\author{Eric M. Monier,}

\affil{Ohio State University, Department of Astronomy,
Columbus, OH 43210}

\and

\author{Jacqueline Bergeron}

\affil{Institut d'Astrophysique de Paris, 98bis Boulevard Arago, F
75014, Paris, France}

\altaffiltext{2}{email: rao@everest.phyast.pitt.edu}

\begin{abstract}

We present optical and near-infrared ground-based imaging results
on four low-redshift damped Ly$\alpha$ (DLA) galaxies. The
corresponding DLA systems were discovered in our {\it Hubble
Space Telescope} spectroscopic surveys for DLA lines in known strong
\ion{Mg}{2} absorption-line  systems toward the quasars B2 0827+243
($z_{DLA}=0.525$), PKS 0952+179 ($z_{DLA}=0.239$), PKS 1127$-$145
($z_{DLA}=0.313$), and PKS 1629+120 ($z_{DLA}=0.532$).  Two of
the four DLA galaxies have confirmed slit redshifts, one has a
photometric redshift consistent with the absorption-line redshift,
and the fourth identification is based on the galaxy's proximity to the
quasar sight line.  The DLA galaxies span a mixture of morphological
types from  patchy, irregular, and low surface brightness to spiral
galaxies. The luminosities range from $0.02L_K^*$ to 
$1.2L_K^*$.  We also discovered several extremely red objects (EROs)
in two of these fields and discuss the possibility that they are
associated with the DLA galaxies.
These observations add to the small but growing list of DLA galaxies 
at low redshift. At the present time, 14 DLA galaxies in the 
redshift range $0.05 \lesssim z \lesssim 1$ have 
been studied. The distributions of DLA galaxy properties for
these 14 cases are discussed and some important trends emerge.
Low-luminosity dwarf galaxies with small impact parameters dominate
this small sample.  Also, four of the five highest column density systems, 
which dominate in the determination of the cosmological neutral gas mass 
density, arise in low surface brightness dwarf galaxies. Zwaan et al. 
have shown that only 15\% of the neutral gas at the present epoch is 
contained in low surface brightness galaxies. Thus, if the low-redshift 
DLA galaxy trends hold up with larger samples, it would indicate that a 
different population of objects is responsible for the
bulk of the neutral hydrogen gas in the universe at $z \approx 0.5$.

\end{abstract}


\section{Introduction}

Surveys for damped Ly$\alpha$ (DLA) absorption lines in the spectra
of quasi-stellar objects (QSOs), or quasars, have been used to study
the distribution of neutral hydrogen in the universe (e.g. Wolfe
et al.  1986; Lanzetta et al. 1991; Rao \& Turnshek 2000, henceforth
RT2000).  These studies indicate that the DLA systems, historically
defined to have  $N_{HI} \ge 2 \times 10^{20}$ atoms cm$^{-2}$, trace
the bulk of the neutral gas mass in the universe up to at least
redshift $z\approx3.5$. At $z>3.5$, inclusion of lower $N_{HI}$
systems in the sub-DLA regime may be necessary to encompass the
bulk of the \ion{H}{1} mass (P\'eroux et al. 2001). In either case, study
of the DLA systems reveals the \ion{H}{1} gas production and consumption
history over a large fraction of the age of the universe. Thus, DLA
galaxies, which are identified by virtue of their \ion{H}{1} gas
cross sections, are the
only population of cosmological objects that simultaneously
reveal information relevant to both star formation and \ion{H}{1} gas
production and consumption, making them crucial 
for understanding the evolutionary history of neutral
gas.  Galaxies selected in optical/IR surveys are
generally only used to track the star formation history of the
universe (see, e.g., Madau, Pozzetti, \& Dickinson 1998). How these two
distinctly selected populations (i.e., \ion{H}{1} cross section-selected 
versus optical/IR-selected) are related is as yet unclear.

Ground-based imaging of high-redshift DLA galaxies has had limited
success, either because the quasar point spread function (PSF) prevents the detection
of objects very close to the quasar sight line, or because the
DLA galaxy is simply too faint, or both. In addition, the faintness of any
candidates close to the quasar sight line  makes it
difficult to obtain confirming redshifts. Imaging of high-redshift
DLA galaxies with the {\it Hubble Space Telescope (HST)} has been
more successful at identifying faint candidates close to the quasar
sight lines, but only a few of these have confirmed redshifts.
See Warren et al. (2001 and references therein)
for details.  The main conclusion of these studies has been that
most high-redshift DLA galaxies are underluminous in comparison to
the Lyman break galaxy population.

Morphologies, colors, and stellar populations of low-redshift
DLA galaxies can be more easily studied, but progress
has been slow mainly because of their rarity.  {\it HST} observations
along many quasar sight lines in traditional spectroscopic survey
mode are needed to find a single DLA at $z<1.65$ (see, e.g., Jannuzi
et al. 1998).  However,
since it is known from optical studies that all high-redshift DLA systems
have accompanying \ion{Mg}{2} absorption, we have to a large extent
circumvented this problem in our {\it HST} surveys by targeting
quasars known to have \ion{Mg}{2} absorption lines in their spectra
(Rao, Turnshek, \& Briggs 1995; 
RT2000).  We reasoned that if this held at high redshift, it would
also be the case at low redshift since galaxies only increase their
metallicities with age. In fact, during the course of this work
we uncovered a new empirical criterion for finding DLA systems.
Approximately half of the systems with \ion{Mg}{2} $\lambda2796$ rest
equivalent width $W_0^{\lambda2796}>0.5$ \AA, as well as \ion{Fe}{2}
$W_0^{\lambda2600}>0.5$ \AA, have  DLA absorption (RT2000), and
the number of DLA systems that do not meet this criterion appears to
be insignificant.  As a result, these targeted surveys have now
nearly tripled the number of DLA systems known at $z<1.65$ to $\approx
30$, which is in comparison to the $\approx 80$ at $z>1.65$ that
have been discovered and confirmed in optical quasar spectra.
Of course, once the strong \ion{Mg}{2}-\ion{Fe}{2} criterion is
used to identify low-redshift DLA candidate systems at $z<1.65$,
{\it HST} UV spectroscopy is still required to obtain a reasonably
precise measurement of $N_{HI}$, which is then used to confirm or
refute the candidate as lying in the DLA regime.  

The few published imaging studies of low-redshift ($z<1.65$) DLA
galaxies have revealed a mix of morphological types. Burbidge
et al. (1996) obtained {\it HST} images along the sight line
towards the BL Lac object AO 0235+165, which contains a $z=0.524$
DLA system selected on the basis of 21 cm absorption; they found a
significant number of galaxies near the sight line, including an 
active galactic nucleus (AGN)
that has broad absorption line (BAL) features and a fainter, late-type 
galaxy. Both of these objects are at the DLA redshift, but  
the late-type galaxy has a smaller impact parameter. 
Le Brun et al. (1997) presented {\it HST} images of six low-redshift
DLA galaxies that included spirals, low surface brightness (LSB)
galaxies, and compact objects, with luminosities ranging from
$0.07L^*$ to $1.4L^*$. Three of these six were selected on the
basis of 21 cm absorption.  In another case, no evidence for a
galaxy at the $z=0.656$ DLA redshift toward the quasar 3C 336
was found, despite very deep ground-based and {\it HST} imaging
(Steidel et al. 1997; Bouch\'e et al. 2001). Also, previously, we
presented results on two DLA galaxies toward the quasar OI 363 at
redshifts $z=0.091$ and $z=0.221$ (Rao \& Turnshek 1998; Turnshek
et al. 2001). The $z=0.091$ galaxy is an LSB galaxy with apparent
spiral structure nearly centered on the quasar sight line, while
the $z=0.221$ galaxy is an early-type dwarf spiral at an impact
parameter of 20 kpc. Here, we present more ground-based imaging
results on three low-redshift DLA systems, which we discovered
in our {\it HST} Cycle 6 survey toward the quasars B2 0827+243,
PKS 0952+179, and PKS 1127$-$145, as well as one new DLA system
which we discovered in our {\it HST} Cycle 9 survey toward the
quasar PKS 1629+120.  We continue to find that the DLA galaxies
are drawn from a mix of morphological types, but, with the increased
total sample size, observed trends are becoming more compelling. In 
the last section of this paper, we summarize
results from imaging studies of a cosmological sample ($z > 0.05$) of
14 DLA galaxies with $z \lesssim 1$ known at the time of publication.

The paper is organized as follows. Details of our observations
are given in \S2. The identification of the DLA galaxies, their
photometric properties, and some details of the fields containing the 
four new low-redshift DLA galaxies are described in \S3.
A discussion and the conclusions are presented in \S4.  All distance
related quantities are calculated using a cosmology with $\Omega_M =
1.0$, $\Omega_\Lambda = 0.0$, and $H_0 = 65$ km s$^{-1}$ Mpc$^{-1}$.
When luminosities are expressed in terms of $L^*$, we assume
$M_B^*=-20.9$ (Marinoni et al. 1999), $M_R^* = -21.2$ (Lin et al. 1996), 
and $M_K^*=-24.5$ (Loveday 2000), where the magnitudes have been 
converted to our adopted cosmology. 

\section{Observations}

The four quasar fields were observed during the period between
1998 December and 2001 June.  Optical images were obtained with
the MDM Observatory 2.4m Hiltner Telescope on Kitt Peak, using the
$1024\times1024$ Templeton CCD (0.285$\arcsec$ pixel$^{-1}$), and with
the 3.5m WIYN telescope on Kitt Peak. Observations with the 
WIYN telescope were made in both classical scheduling
mode, using the Mini-Mosaic $4096\times2048$ SITe CCD pair
(0.141$\arcsec$ pixel$^{-1}$), and queue mode, by the WIYN queue
observing team using a $1024\times1024$ Tektronix CCD (0.195$\arcsec$
pixel$^{-1}$).  The near-infrared images were obtained at the 3.0m
NASA IRTF on Mauna Kea, using NSFCAM in combination with a $256\times256$
InSb detector array (0.30$\arcsec$ pixel$^{-1}$), and the 3.6m
ESO NTT on La Silla, using SOFI, which uses a $1024\times1024$
HgCdTe detector array (0.292$\arcsec$ pixel$^{-1}$).  The images
were processed using the recommended procedures, and standard-star
observations were used to calibrate the photometry.  The observations
are summarized in Table 1. We note that some of the observations
listed in Table 1 were taken under less than ideal conditions, but
were still useful for confirmation of some of the derived results.
Some of the images were smoothed to enhance LSB
features, and these are shown in \S3. The smoothing was performed 
using a Gaussian with $\sigma \approx 1$ pixel. On average, this degraded 
the seeing by 20\%. The smoothed images were used to delineate the 
LSB features and set the aperture for doing the photometry, but 
photometric measurements were made on the unsmoothed images. Faint features
in the smoothed images were deemed real if they were detected in at least 
two filters at greater than a $2\sigma$ level of significance. 
Limiting $3\sigma$ surface brightnesses and seeing measurements 
for the final combined (unsmoothed) images used in the analysis presented here
are given in Table 2.

\begin{deluxetable*}{ccclc}
\tabletypesize{\footnotesize}
\tablewidth{0pt}  
\tablecaption{Journal of Imaging Observations}  
\tablehead{ 
\colhead{QSO Field}  & 
\colhead{Filter} &
\colhead{Telescope}  & 
\colhead{Observation}  & 
\colhead{Exp. Time}\\[.1ex]  
\colhead{}  & 
\colhead{}  & 
\colhead{}  & 
\colhead{Dates}  &
\colhead{(m)}  }  
\startdata  
B2 0827+243         & U       & MDM        & 1999 Nov 13,14 & 100 \\  
\nodata             & B       & WIYN       & 1999 Jan 18 & 54 \\  
\nodata             & R       & WIYN       & 1999 Jan 18 & 15 \\  
\nodata             & I       & MDM        & 1999 Nov 14 & 28 \\  
\nodata             & K       & IRTF       & 1999 Apr 27 & 62 \\ 
PKS 0952+179        & U       & MDM        & 1999 Feb 17$-$20 & 180 \\
\nodata             & B       & MDM        & 1999 Feb 18,20 & 60 \\
\nodata             & R       & MDM        & 1999 Feb 19,20 & 75 \\
\nodata             & R       & WIYN       & 1999 Mar 20 & 27 \\
\nodata             & J       & NTT        & 1999 Jan 02 & 60 \\
\nodata             & K$_s$   & NTT        & 1999 Jan 02 & 60 \\
\nodata             & K       & IRTF       & 1998 Dec 12 & 20 \\
\nodata             & K       & IRTF       & 2000 Mar 06 & 63 \\  
PKS 1127$-$145      & U       & MDM        & 1999 Feb 18,19 & 120 \\  
\nodata             & B       & MDM        & 1999 Feb 19,20 & 60 \\  
\nodata             & R       & MDM        & 1999 Feb 18$-$20 & 80 \\
\nodata             & I       & MDM        & 1999 Nov 13 & 14 \\ 
\nodata             & I       & MDM        & 2001 Feb 26 & 30 \\
\nodata             & J       & NTT        & 1999 Jan 02 & 80 \\  
\nodata             & K       & IRTF       & 1998 Dec 13 & 60 \\
\nodata             & K       & IRTF       & 1999 Apr 28 & 63 \\
PKS 1629+120        & U       & WIYN       & 2001 Jun 25,27 & 60 \\
\nodata             & B       & MDM        & 2000 Sep 29 & 45 \\
\nodata             & R       & MDM        & 2000 Sep 29,30 & 45 \\
\nodata             & J       & IRTF       & 2000 Sep 30 & 50 \\
\nodata             & K       & IRTF       & 2000 Sep 29 & 60 \\
\enddata
\end{deluxetable*}

\section{DLA Galaxy Identification and Photometry}
 
As detailed below, two of the DLA galaxies have confirmed slit redshifts, 
one has a photometric redshift consistent with the DLA absorption redshift, 
and one is identified based on its proximity to the 
quasar sight line. It should be kept in mind that the degree of confidence 
for any ``identification'' of a DLA galaxy is variable, with the confidence
being highest when the candidate DLA galaxy has a low impact parameter
to the QSO sight line {\it and} there is a confirming slit spectrum
showing that it is at the DLA system redshift.  However, even when the
confidence is relatively high, it is possible that a 
fainter galaxy or a galaxy with a smaller impact parameter,
might be the actual DLA absorber. We proceed with the assumption that
a neutral gas cloud (or clouds along the line of sight) 
associated with the galaxy we identify as the DLA galaxy is the site 
of the DLA absorption. 

When possible, bright stars in the field were used to model the quasar
PSF, which was then subtracted before any
photometric measurements were made.  This was sometimes not possible in the
IRTF images since there were no bright stars in the small field of
view (77\arcsec) and no PSF star was observed separately.

\subsection{B2 0827+243}

This quasar sight line contains a DLA system at $z=0.525$ with column
density $N_{HI}=(2.0\pm0.2)\times10^{20}$ atoms cm$^{-2}$ (RT2000).
Figure 1 shows $\approx$ 30\arcsec\ $\times$ 30\arcsec\ $B$, $R$, $I$, and $K$ 
images of this field. The quasar PSFs have 
been subtracted and the residuals
masked out. The  galaxy 6\arcsec\  to the east of the sight line,
labeled G1, is identified as the DLA galaxy.  Steidel et al. (2002)
report a spectroscopic redshift of  $z=0.5258$  for G1.
They also present an {\it HST}-WFPC2 F702W image of this
field and note the possible presence of a satellite galaxy 
$\approx 2$\arcsec\ west-northwest of the center of G1. We 
 a posteriori found a  hint of a southwestern extension to G1 in our
smoothed $I$-band image (see inset in Fig. 1). Of course, our $I$-band 
image was obtained in 1.1\arcsec\ seeing, compared to the $\approx0.1$\arcsec\
seeing of the {\it HST} image. Thus, the satellite noted by
Steidel et al. (2002) is unresolved and is not distinct from G1 in our 
ground-based image. Instead, it is possible that the southwestern extension 
we detect is an LSB feature of the interacting system
and was not detected in the {\it HST} image. In either case, the imaging data
are consistent with G1 being a disturbed spiral galaxy.
Steidel et al. (2002) also present Keck spectra showing the 
kinematic properties of  G1 and of the $z=0.525$ \ion{Mg}{2} absorption-line 
system. They note that the satellite galaxy might be responsible
for the apparent reversal in the direction of the radial velocity profile 
of G1 seen at the west end. This interaction might also be dispersing the 
gas out to large galactocentric radii,  leading to the 
high \ion{H}{1} column density of the absorbing gas and to the 
$\approx 270$ km s$^{-1}$ wide, four-component, \ion{Mg}{2} absorption line.
The unsaturated \ion{Mg}{1} $\lambda 2852$
 line shows similar structure, with the central 
two components that are $\approx 50$ km s$^{-1}$ apart being the strongest.
 The 21 cm absorption line has been measured at moderate spectral
resolution and covers $\approx 50$ km s$^{-1}$, although it is
uncertain how many absorbing components might be contributing to the
profile (Kanekar \& Chengalur 2001a).  Based on a comparison of
redshift and velocity spread, it is likely that the \ion{H}{1} absorption
arises in gas associated with the two main \ion{Mg}{1} absorbing clouds.
 Indeed, there is compelling evidence
that strong \ion{Mg}{2} absorbers that show 21 cm absorption are also 
associated with strong \ion{Mg}{1} absorption (Lane 2000). Moreover, 
systems with larger \ion{Mg}{1} $\lambda2852$ rest equivalent width (or velocity
spread when observed at high resolution) also have larger 21 cm line 
velocity spreads (Lane 2000). 

Although the impact
parameter of G1 is large ($b=34$ kpc), its high inclination angle
should increase the likelihood that DLA column densities could be observed 
at large galactocentric distances.  But it is also possible that an
object  hidden under the quasar PSF might be the actual absorber. To
explore this possibility, we subtracted the quasar PSF in the WFPC2 image
using the procedure described in Hamilton, Casertano, \& Turnshek
(2002), but found no convincing evidence for a  galaxy hidden under
the quasar PSF. 
\begin{figure*}
\centerline{\epsfig{file=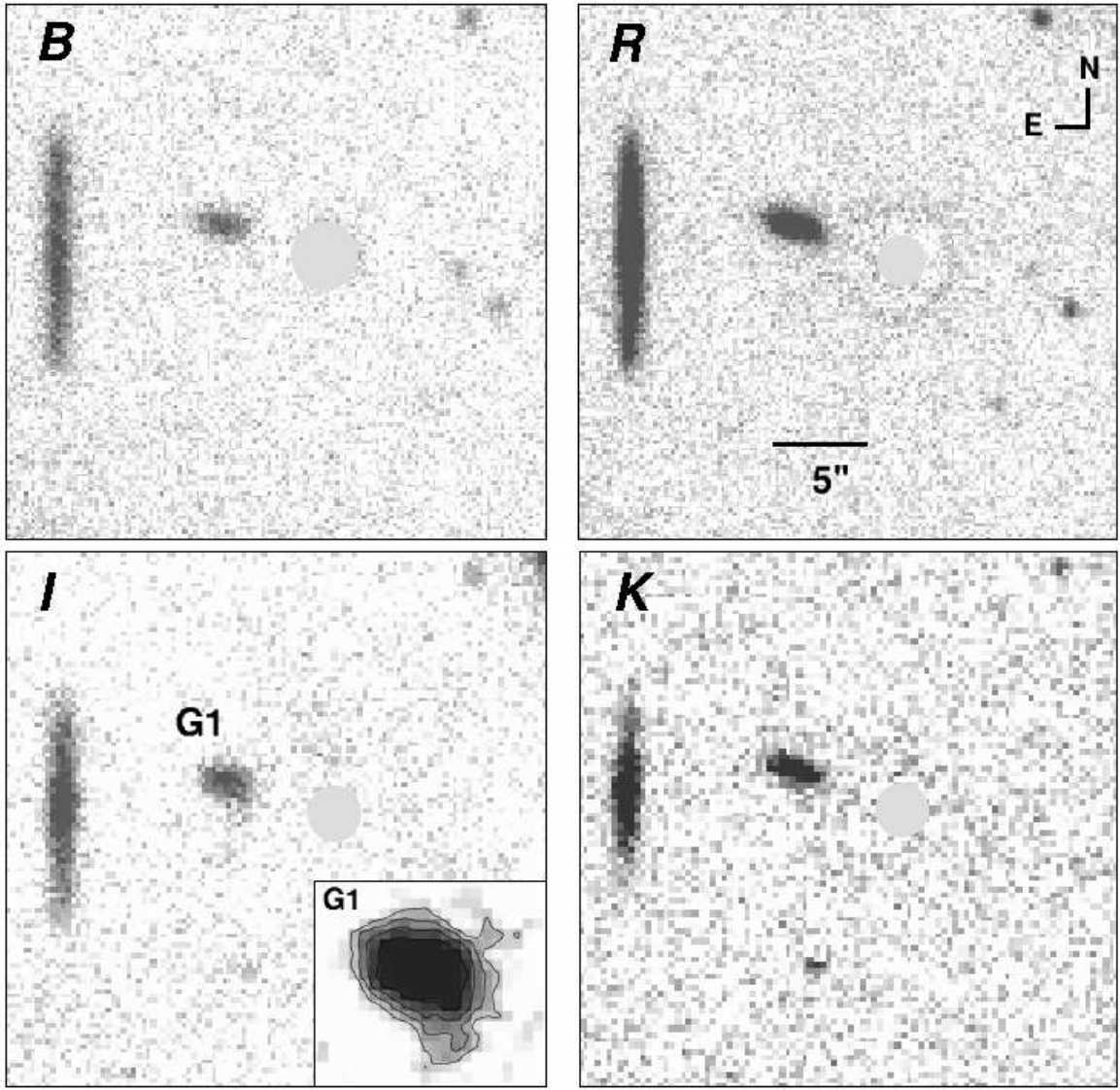, width=6in}}
\caption{\footnotesize $B$, $R$, $I$, and $K$ images of the B2 0827+243 field that
contains a DLA system at $z=0.525$. Quasar PSFs have been
subtracted, and the residuals have been masked. G1 has a measured
redshift of $z=0.5258$ (Steidel et al. 2002) and is identified as the
DLA galaxy. The inset is a smoothed image of G1 in $I$, where a gaussian 
smoothing with $\sigma=0.8$ pixels has been applied. The outermost contour is 
$1\sigma$ above the sky background. 
 Note that the north-south--oriented edge-on galaxy $\approx 15$\arcsec\
east of the quasar is at $z=0.199$ (Steidel et al. 2002).}
\end{figure*} 

\begin{center}
\begin{deluxetable*}{lccccccccccccc}
\tablewidth{0pt}  
\tablecaption{$3\sigma$ Surface Brightness Limits ($\mu$) and Seeing Measurements
($\Theta$\tablenotemark{a} )}  
\tablehead{ 
\colhead{Field}   & 
\colhead{$\mu_U$} &
\colhead{$\mu_B$} & 
\colhead{$\mu_R$} & 
\colhead{$\mu_I$} &
\colhead{$\mu_J$} & 
\colhead{$\mu_K$} &
\colhead{} &
\colhead{$\Theta_U$} & 
\colhead{$\Theta_B$} & 
\colhead{$\Theta_R$} & 
\colhead{$\Theta_I$} & 
\colhead{$\Theta_J$} & 
\colhead{$\Theta_K$}  \\[.2ex] \cline{2-7} \cline{9-14}
\colhead{} &
\multicolumn{6}{c}{(mags/arcsec$^2$)} &
\colhead{} &
\multicolumn{6}{c}{(arcsec)}
}
\startdata 
B2 0827+243    & 24.0 & 25.1 & 24.1 & 22.9    &\nodata & 19.6 &
& 1.3 & 1.0 & 0.8 & 1.1 & \nodata & 0.7 \\
PKS 0952+179   & 24.2 & 24.8 & 24.9 & \nodata & 22.2   & 20.8 &
& 1.2 & 1.1 & 0.7 & \nodata & 0.7 & 0.9 \\ 
PKS 1127$-$145 & 24.0 & 25.0 & 24.5 & \nodata & 22.8   & 20.4 &
& 0.9 & 1.2 & 0.9 & \nodata & 0.7 & 0.8 \\ 
PKS 1629+120   & 23.6 & 24.6 & 24.2 & \nodata & 21.1   & 20.4 & 
& 1.3 & 1.3 & 1.2 & \nodata & 1.0 & 1.0 \\ 
\enddata
\tablenotetext{a}{FWHM}
\end{deluxetable*}
\end{center}

\subsubsection{Photometry of G1}

Photometric measurements of G1 along with $1\sigma$ uncertainties
are given in Table 3. Results from stellar population spectral evolutionary synthesis 
model fits redshifted to $z=0.525$ are shown in Figure 2.  Details of
the models and fitting  procedures are described in the Appendix.  The
best-fit single-burst model, a 0.05 Gyr old burst with $E(B-V)=0.6$,
results in a reduced $\chi^2$ of 5.5.  A family of two-burst models having
approximately equal combinations by mass of (1) a dusty
$\left[0.6\lesssim E(B-V)\lesssim 0.9\right]$, young (0.01 Gyr) burst 
and (2) a nearly dust free [$0.0\lesssim E(B-V)\lesssim 0.2$], 0.2$-$0.6 Gyr 
old burst have the smallest reduced
$\chi^2$ values, but these are  also statistically poor fits
with reduced $\chi^2$ values between 4.8 and 5.2.  The fit with the
smallest reduced $\chi^2$ is shown in Figure 2.
Apparent magnitudes from Table 3 along with 
$K$-corrections derived from this best-fit model give absolute magnitudes
for G1 of $M_U=-20.3$, $M_B=-20.6$, $M_R=-21.7$,
$M_I=-22.1$ and $M_K=-24.7$ at $z=0.525$. 
\begin{figure}
\plotone{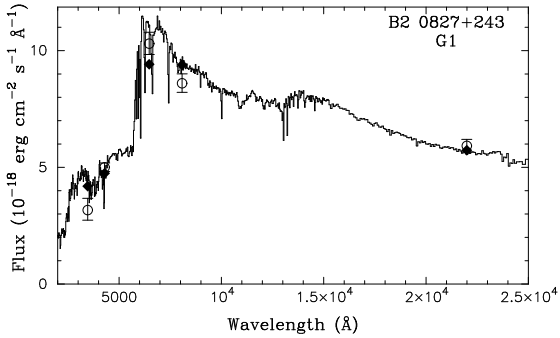}
\caption{\footnotesize Best-fit two-burst model of a galaxy at $z=0.525$ 
overlaid on the photometry of G1  in
the B2 0827+243 field. The circles with $1\sigma$ error bars represent
photometric measurements. For comparison, the filled diamonds represent model 
flux values determined at the effective wavelength of the filter 
by convolving the filter response function with the 
model spectral energy distribution. The model is a two-burst combination of 
(1) a 48\% by mass burst that is young (0.01 Gyr) and dusty, with $E(B-V)=0.6$,
and (2) a 52\% by mass burst that is older (0.6 Gyr) and dust 
free, with $E(B-V)=0.0$. }
\end{figure}

\begin{center}
\begin{scriptsize}
\begin{deluxetable*}{lcrrcccccc}
\tablewidth{0pt}  
\tablecaption{Object Photometry\tablenotemark{a}}
\tablehead{ 
\colhead{Field} &
\colhead{Object} &
\colhead{$\Delta\alpha$\tablenotemark{b}}  & 
\colhead{$\Delta\delta$\tablenotemark{b}}  &
\colhead{$m_U$} & 
\colhead{$m_B$} & 
\colhead{$m_R$} & 
\colhead{$m_I$} &
\colhead{$m_J$} &
\colhead{$m_K$} \\[.2ex] 
\colhead{} & 
\colhead{} &  
\colhead{\arcsec} &
\colhead{\arcsec} &  
\colhead{} &
\colhead{} &
\colhead{} &
\colhead{} &
\colhead{} &
\colhead{} } 
\startdata 
B2 0827+243 & G1 &+5.8  & +1.9 &  22.61 (0.16) & 22.76 (0.04) & 20.82 (0.05) & 20.34 (0.05) & \nodata &17.07 (0.05) \\
PKS 0952+179 & 1  &+2.9   &0.0    & \nodata  & \nodata    & \nodata    & \nodata  & 22.3 (0.2) & 21.2 (0.3) \\   
&2  &$-$1.4 &$-$1.8 & \nodata  & \nodata    &\nodata     & \nodata  & 23.1 (0.4) & 20.8 (0.1) \\  
&3  &+3.7   &$-$4.3 & \nodata  & 25.9 (0.3) & 25.2 (0.2) & \nodata  & 22.5 (0.3) & 20.7 (0.2) \\  
&4  &$-$3.5 &+5.1   & \nodata  & 25.0 (0.2) & 24.5 (0.2) & \nodata  & 22.1 (0.2) & 20.1 (0.2) \\  
&5  &+6.7   &$-$4.4 & \nodata  & 24.8 (0.2) & 24.0 (0.1) & \nodata  & 23.2 (0.4) & 20.4 (0.2) \\  
&6  &$-$6.8 &+0.5   & \nodata  & 27.1 (0.7) & 27.1 (0.8) & \nodata  & 22.3 (0.3) & 20.6 (0.2) \\
&7  &$-$7.5 &$-$2.0 & \nodata  & \nodata    & 25.9 (0.3) & \nodata  & 22.0 (0.3) & 19.4 (0.1) \\  
&8  &$-$7.0 &$-$3.9 & \nodata  & 27.9 (1.6) & 25.1 (0.2) & \nodata  & 23.0 (0.4) & 21.8 (0.4) \\  
&9  &$-$1.9 &+8.4   & \nodata  & 24.9 (0.2) & 24.5 (0.1) & \nodata  & 22.4 (0.3) & 21.0 (0.2) \\  
&10 &$-$7.8 &$-$8.7 & \nodata  & 25.0 (0.2) & 24.5 (0.2) & \nodata  & 22.5 (0.3) & 21.2 (0.3) \\
PKS 1127$-$145 & 1  & $-$3.8 & +0.3   & 23.00 (0.19) & 24.09 (0.15) & 22.58 (0.08) & \nodata & 21.87 (0.13) & 20.64 (0.15) \\ 
&2  & $-$2.5 & $-$5.8 & $>$23.9      & $>$24.8      & $>$24.5      & \nodata & 22.02 (0.13) & $>$20.3      \\  
&3  & +5.4   & $-$1.6 & 23.55 (0.21) & 23.68 (0.09) & 22.40 (0.07) & \nodata & 20.94 (0.07) & 20.23 (0.20) \\ 
&4  & $-$3.2 & +7.2   & 23.93 (0.25) & 24.94 (0.17) & 24.11 (0.19) & \nodata & 22.93 (0.22) & 20.96 (0.30) \\
PKS 1629+120 & G1 & +0.8 &$-$2.9 & 22.61 (0.09) & 23.15 (0.11) & 21.37 (0.04) & \nodata & 19.20 (0.07) & 17.65 (0.06) \\
\enddata
\tablenotetext{a}{Numbers in parentheses are $1\sigma$ errors.}
\tablenotetext{b}{Relative to QSO.}
\end{deluxetable*}
\end{scriptsize}
\end{center}

\subsection{PKS 0952+179}

The DLA system towards this quasar is at redshift $z=0.239$,  with
column density $N_{HI}=(2.1\pm0.3)\times10^{21}$ atoms cm$^{-2}$
(RT2000).  Figure 3 shows 20\arcsec\ $\times$ 20\arcsec\ images of
this  field through the $B$, $R$, $J$, and $K$ filters. The quasar PSF
has been subtracted in the $B$, $R$, and $J$ images and the  residuals
have been masked out. The quasar has simply been masked out in the $K$-band 
image because there were no suitable bright stars in the
IRTF-NSFCAM field of view that could be used to perform a good PSF
subtraction of the quasar light. All four images have been smoothed to enhance
LSB features.  Objects identified in the $J$ image
that are detected in at least one of the $B$, $R$, and $K$ images have been
numbered sequentially in order of increasing radial  distance from the
quasar. The objects labeled 1 and 2 are detected in both $J$ and $K$
and, because of their proximity to the quasar sight line, are our best
candidates for  the DLA galaxy.  Their morphology suggests that they
might be two  nearly edge-on galaxies. While the $J$ image, with a
seeing of  0.7\arcsec\ and limiting surface brightness of 24
mag/arcsec$^2$, is exceptionally good by ground-based imaging
standards,  higher resolution imaging, for example with NICMOS on 
{\it HST}, is required to be better able to resolve these features  and,
perhaps, comment on the  nature of the $z=1.478$ quasar host  galaxy
as well (see Kukula et al. 2001).  Table 3 gives the positions
of the labeled objects relative to the quasar along with 
 apparent magnitudes and $1\sigma$ uncertainties.

\begin{figure*}
\centerline{\epsfig{file=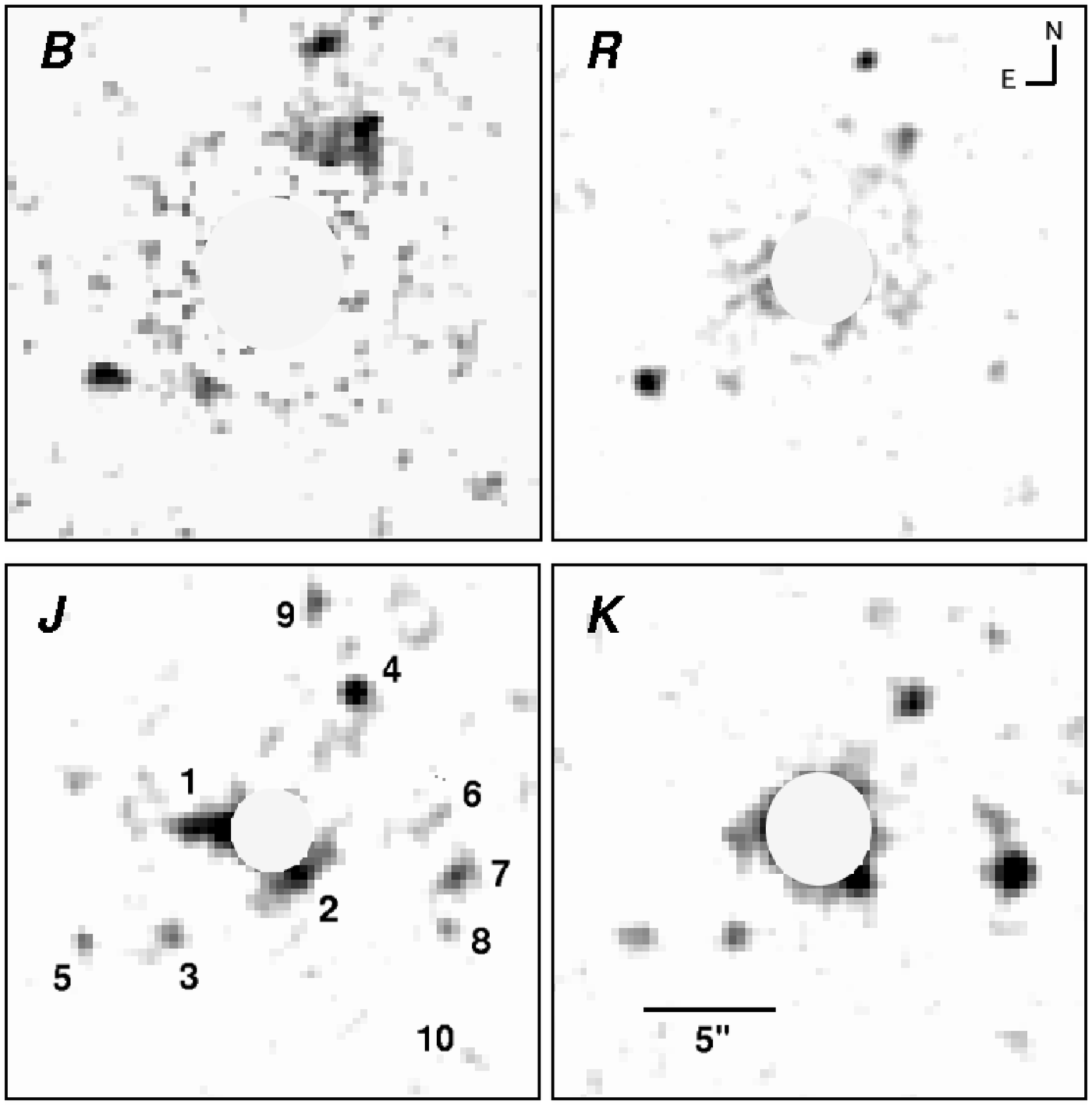, width=6in}}
\caption{\footnotesize Smoothed $B$, $R$, $J$, and $K$ images of the PKS 0952+179
field that contains a DLA system at $z=0.239$.  The PSF of the
quasar has been subtracted in the $B$, $R$, and $J$ images, but not in
the $K$ image as there was no suitable PSF star in the field. The
residuals in $B$, $R$, and $J$ and the quasar in $K$ have been masked.
Objects are labeled only on the $J$ image, for clarity. All 10
of these are detected in $K$. Features in the $B$ and $R$ 
images that lie at the edges of the  mask are  artifacts of the PSF 
subtraction process; only objects labeled 3, 4, 5, 8, 9, 
and 10 are detected in  $B$, while those labeled 
3, 4, 5, 7, 8, 9, and 10 are detected in $R$. Objects 1 and 2 are the 
best candidates for the DLA
galaxy. Object 7 is an ERO, with $R-K=6.5$. See \S3.2 for more details.}
\end{figure*}

\subsubsection{An ERO at $z=0.239$?}

The brightest of these in $K$, object 7, is an extremely red
object (ERO) with  $R-K=6.5$ (see Table 3).  EROs are generally
thought to either be elliptical galaxies at $z\gtrsim1$ whose red colors are
due to large $K$-corrections or star forming galaxies whose red colors
are due to heavily obscured stellar or AGN emission (e.g. Cimatti et
al. 1999; Dey et al. 1999; Daddi et al. 2000; Smith et al. 2001). The
lowest redshift ERO known is the $z=0.65$ galaxy, PDF J011423 (Afonso
et al. 2001), whose red colors  and spectral energy distribution (SED) are
consistent with its being a dusty  starburst with 5$-$6 magnitudes of
optical extinction.  Other EROs classified as starbursts
have similar dust extinctions (Afonso et al. 2001).  PDF J011423 is
bright in $K$ ($m_K=15.3$) and has extremely red colors with $R-K=5.8$
and $J-K=3.1$. In comparison, object 7 is relatively faint with
$K=19.4$, and is also extremely red, with $R-K=6.5$ and $J-K=2.6$. The
redshift of object 7 is not known, but if it is at the DLA redshift,
it would have to be a star forming region that is heavily obscured by
dust.  At $z=0.239$, $K=19.4$ corresponds to $M_K=-20.7$, where a $K$-band 
$K$-correction of $-0.2$ mag has been applied (Cowie et
al. 1994). This implies an absolute luminosity of $L_K=0.03L_K^*$,
comparable to the luminosity of a single star-forming region within a 
galaxy. We find that, based on G. Bruzual \& S. Charlot (2003, in preparation) 
GISSEL99 galaxy spectral evolutionary
synthesis models, its colors can be attributed to a region of star
formation with age $\lesssim 0.6$ Gyr at $z=0.239$, if the extinction is
$A_V \gtrsim 4.8$.  Thus, the luminosity, colors, and implied
extinction are consistent with object 7 being at $z=0.239$. If
confirmed, it would be the lowest  redshift ERO known.

However, the observational constraints are not tight enough to rule
out the possibility that object 7 is a higher redshift elliptical galaxy.
Its colors fall on the ``starburst'' side, but near the edge, of the
$R-K$ versus $J-K$ plane used to classify EROs (Pozzetti \& Mannucci
2000).  Uncertainties in our photometry, as well as in the definition
of the plane, do not  preclude object 7 from crossing the plane and being
classified as a passively-evolving elliptical galaxy at $z\approx1.5$
(Cimatti et al. 1999).  In this case, object 7 might be part of  a cluster
of galaxies that also includes  the quasar PKS 0952+179
($z_{em}=1.478$); it would be $\sim L^*$ at this redshift.  We detect
a significant number of EROs in the extended 5\arcmin$\times$5\arcmin\
field surrounding this quasar (D. B. Nestor et al. 2003, in preparation),
and this might be supporting evidence for the overdensity of EROs
found by Cimatti et al. (2000) around radio-loud quasars. Thus, the
proximity of object 7 to the DLA galaxy might just be a coincidence.

\subsubsection{The DLA Galaxy}
   
While the nature of none of the objects labeled in Figure 3 is known for
certain, it is of interest to examine the possibility that they are
star-forming regions associated with the DLA galaxy. The colors of all
the labeled features are consistent with them being at $z=0.239$, but
with varying degrees of obscuration. The implications might be that
the DLA galaxy is either a patchy LSB galaxy or a disturbed system in
which object 7 is a recently triggered star-forming region. The galaxy
would extend anywhere from $\approx 45$ kpc (objects 1$-$7) to
$\approx 55$ kpc (objects 5$-$7).  
The total luminosity of the PKS 0952+179 DLA galaxy cannot
be  measured, since some of it is likely to be hidden by the quasar
PSF. The total luminosity of all the objects labeled in Figure 3 is
$m_K = 18.1$. If this is considered a lower limit to the luminosity of
the DLA galaxy, then $M_K < -22.0$ or $L_K > 0.1L_K^*$.  If only 
objects 1 and 2 were part of the DLA galaxy, it would extend $\approx 24$
kpc, with $m_K < 20.2$, $M_K < -19.5$, and  $L_K >0.01L_K^*$. 
If objects 1 and 2 were two edge-on galaxies, as their $J$-band 
morphologies might suggest, then possibly half of the galaxy that is object 1 
is obscured by the quasar PSF. In this case,
each of the edge-on galaxies would be a dwarf with a luminosity on 
the order of $0.01L_K^*$. 

However, in the absence of higher resolution imaging, we tentatively 
identify the DLA galaxy as objects 1 and 2, since these have the smallest impact
parameters, and since there is no observable emission between these and 
objects 3$-$10.  Given the assumption that half of object 1 is obscured by the
quasar PSF, objects 1 and 2 have a total luminosity of $L_K = 0.02L_K^*$. 
Since these objects extend into the quasar PSF, their impact parameter 
cannot be determined to better than the radius of the circle that encloses 
all residuals left over from subtracting the quasar PSF in the 
unsmoothed $J$-band image. This is because subtracting the PSF resulted in 
noisy residuals near the PSF core, and no useful information could be extracted
within this radius. The radius was
measured to be 1.2\arcsec\ which, at $z=0.239$, implies $b<4.5$ kpc.

We note that luminous galaxies in the extended field surrounding this
quasar have spectroscopic redshifts different from the DLA absorption
redshift (Bergeron \& Boiss\'e 1991) and that the $z=0.239$ system is
the only known absorption-line system in the spectrum of this quasar.
High resolution observations of the $z=0.239$ \ion{Mg}{2} absorption 
line do not exist, and so the kinematic structure of the \ion{Mg}{2}-absorbing
gas has not been studied in detail.  However, the 21 cm absorption line 
is narrow with a FWHM of only 7.7 km s$^{-1}$ (Kanekar \& Chengalur 2001a), 
indicative of simple kinematic structure.  

\subsection{PKS 1127$-$145}

\begin{figure*}
\centerline{\epsfig{file=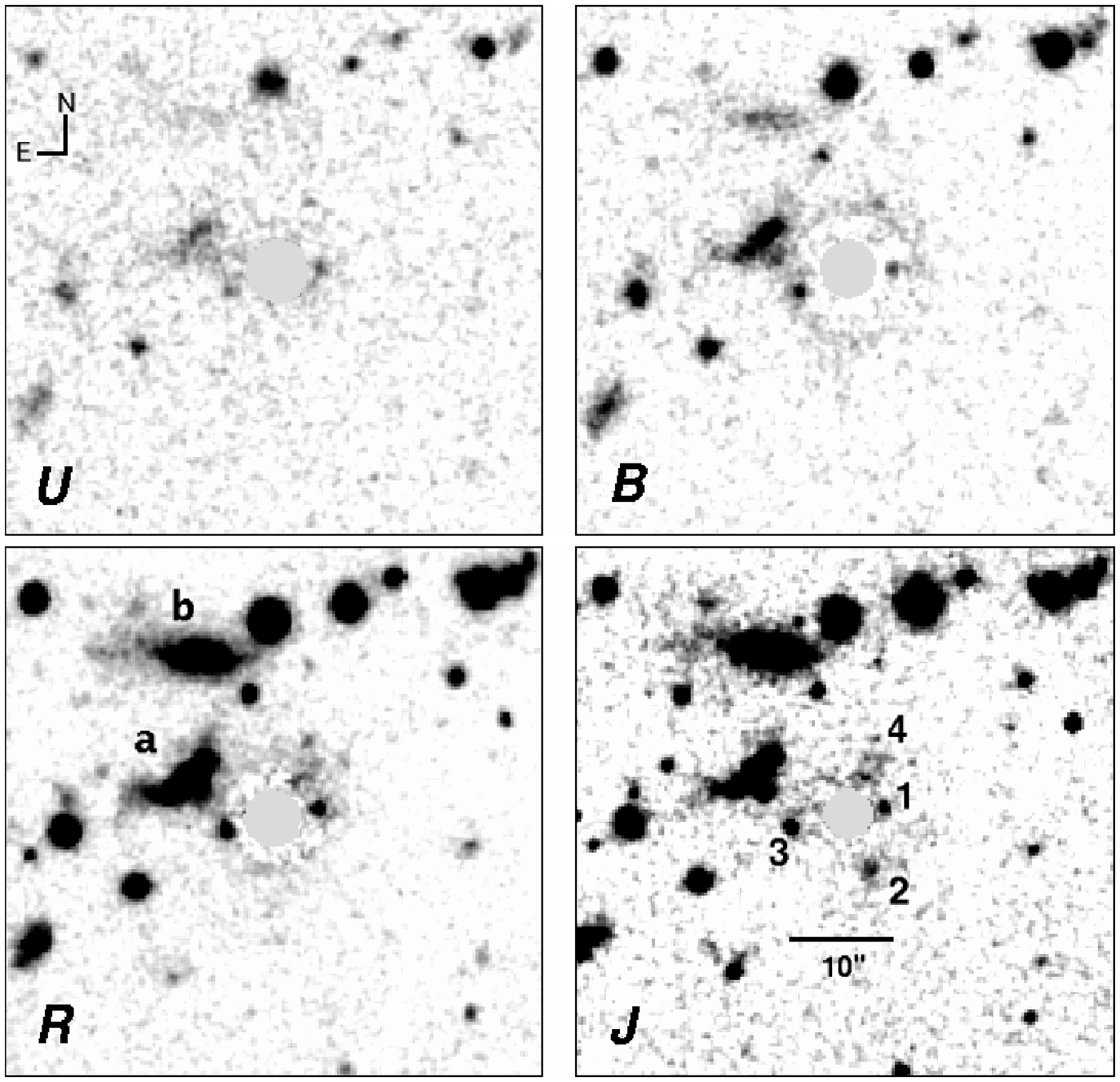, width=6in}}
\caption{\footnotesize Smoothed $U$, $B$, $R$, and $J$ images of the PKS 
1127$-$145 field that contains a DLA system at $z=0.313$.  The PSFs of the
quasar have been subtracted and the residuals masked. The DLA galaxy is
assumed to be the patchy/LSB structure that extends between objects 3 and
1 in the east-west direction and between objects 4 and 2 in the north-south
direction. Object 1 has a measured slit redshift of $z=0.3121$.  The two
large spirals in the field, labeled ``a'' and ``b''  are also  at the
DLA redshift. Note the presence of several EROs that are visible in $J$ but 
are faint or below the detection limit in $R$. The ringlike structures 
around the quasar in $B$ and $R$, left
over from the quasar PSF subtraction process,  appear enhanced because of the
smoothing and are not real.}
\end{figure*}
This  DLA system is at $z=0.313$ with column density
$N_{HI}=(5.1\pm0.9)\times10^{21}$ atoms cm$^{-2}$ (RT2000).  Figure 4
shows $\approx 50\arcsec \times 50\arcsec$ images of this field in
$U$, $B$, $R$, and $J$. The quasar PSF has been subtracted in each of the
four images and the residuals have been masked out. Furthermore, the
images have been smoothed to enhance LSB
features.  The two large spiral galaxies, labeled
``a'' and ``b'', are at the DLA redshift (Bergeron \&  Boiss\'e 1991).
Galaxy a is clearly warped and has a dwarf companion
$\approx2$\arcsec\ to  the southwest.  Galaxy b has a  faint extension 
towards the east that is visible  in $R$ and $J$.  The object 3\arcsec\  
west of the quasar, labeled ``1,'' is an emission-line object with a confirmed 
redshift of $z=0.3121\pm 0.0003$ (Lane et al.  1998). No emission 
lines were detected in the spectrum of object 3, and so its redshift
is not known. However, the $J$ band image in  Figure 4 suggests that there 
may be LSB features to the east of the quasar that 
extend out to the position of object 3. Thus, we identify the DLA galaxy 
as the patchy/irregular LSB structure visible primarily
in $J$  that extends $\approx 10\arcsec$ (44 kpc) in the N-S
as well as E-W directions and encompasses objects 1, 2, 3, and 4.
\footnote{We note that objects in this field were discussed in 
Lane (2000) where galaxies a and b were labeled G1 and G2, and 
objects 1 and 3 were labeled G3 and G4, respectively.}

Since the DLA galaxy as defined above overlaps with the quasar PSF, its impact 
parameter cannot be determined to better than the radius of the circle that encloses 
all residuals left over from subtracting the quasar PSF in the 
unsmoothed $J$-band image.  This radius, which we assume is an upper limit
to the DLA galaxy's impact parameter, was measured to be 
1.5\arcsec, which, at $z=0.313$, implies $b<6.5$ kpc. 
No useful information could be extracted within this radius. The DLA galaxy is 
quite possibly the remains of a dwarf galaxy (see \S3.3.1) that is being tidally 
disrupted by the more massive spirals, and in which these four objects represent 
regions of recent star formation.

\subsubsection{Photometry}

Photometric measurements of objects 1, 2, 3, and 4
are given in Table 3 along with their $1\sigma$ uncertainties. 
Object 2 is below the detection limit in $R$ and has
$R-J>2.3$. Lower limits to the $U$, $B$, $R$, and $K$ magnitudes of
object 2 were measured with the same aperture size used to measure its
magnitude in $J$. 

Objects 1, 3, and 4 have blue colors. Stellar population synthesis model 
fits to the photometry of these star-forming regions are shown in Figure 5. 
Details of the models are given in the Appendix.  For object 1, a family of
single bursts fits the data well. These have burst ages ranging from 0.1 
to 0.3 Gyr and $0.0\lesssim E(B-V)\lesssim 0.10$. The top panel of Figure 5 
shows the best-fit model, which is a 0.2 Gyr old burst with no extinction.
This fit resulted in a reduced $\chi^2$ of 1.37.

\begin{figure}
\epsscale{1.0}
\plotone{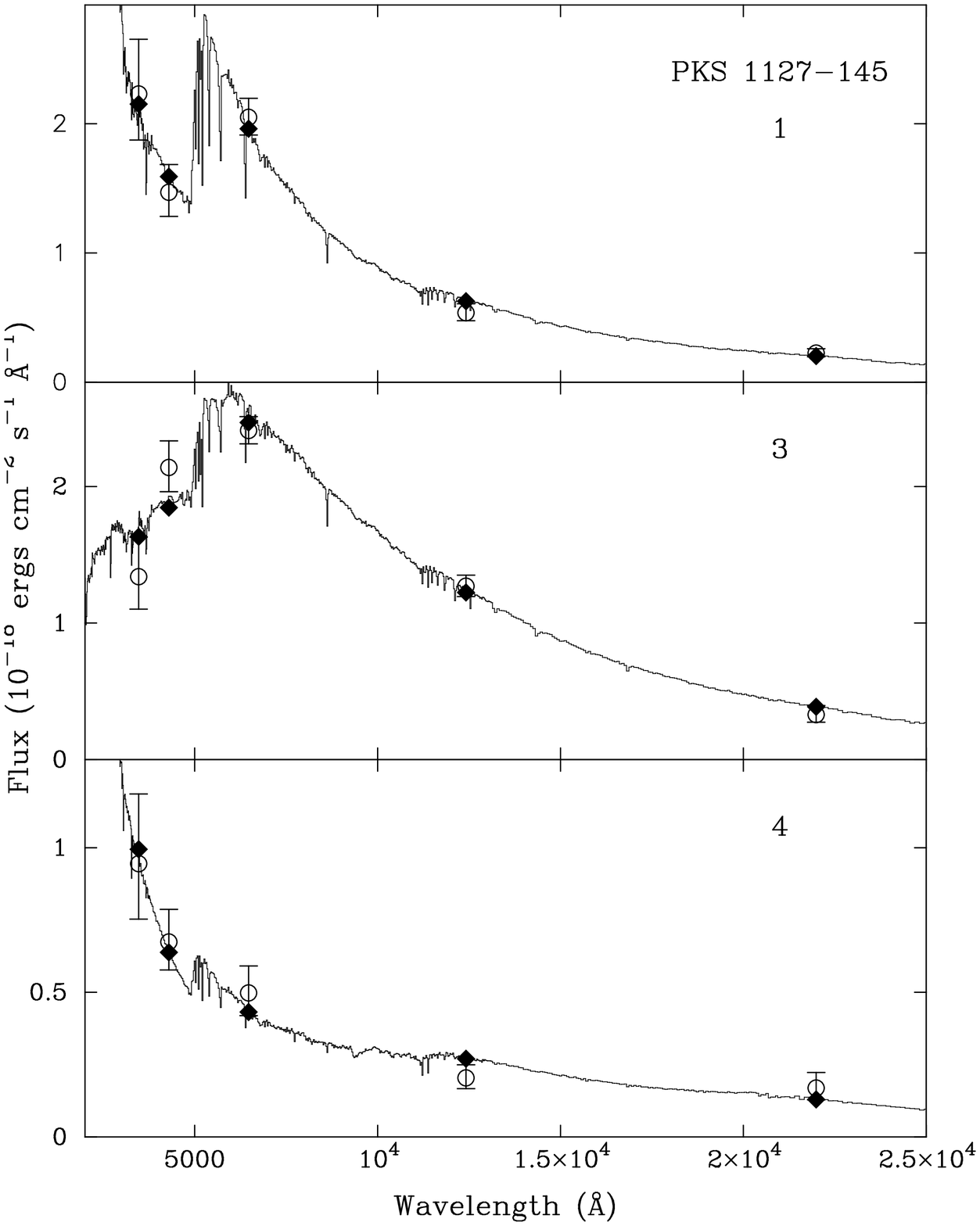}
\caption{\footnotesize Stellar population spectral synthesis model fits at $z=0.313$
for objects 1, 3, and 4 in the PKS 1127$-$145 field. Symbols are as in Fig. 2. The 
best-fit model for object 1 
is a single 0.2 Gyr old burst with no extinction. Object 3 is best fitted by a two-burst
model with (1) an $\approx 70$\% by mass burst that is young (0.001 Gyr) and dusty, with 
$E(B-V)\approx 0.7$, and (2) an $\approx 30$\% by mass burst that is 1.0 Gyr old 
with no dust. Object 4 is best fitted by a single 0.01 Gyr burst with $E(B-V)=0.2$.}
\end{figure}

Single burst models are not a good fit to the photometry of object 3.
The best fit has a high reduced $\chi^2$ of 3.80.  For two-burst models the best 
fit, with a reduced $\chi^2$ of 1.77, is obtained when the populations are
(1) an $\approx 70$\% by mass burst that is young (0.001 Gyr) and dusty, with 
$E(B-V)\approx 0.7$, and (2) an $\approx 30$\% by mass burst that is 1.0 Gyr old 
with no dust. This two-burst model is shown in the middle panel of Figure 5.

Object 4 is well fitted by a family of single bursts having ages in the
range of 0.01$-$0.05 Gyrs and $0.05 \lesssim E(B-V)\lesssim 0.2$.  
Of these, the best-fit model is a 0.01 Gyr burst  with $E(B-V)=0.2$ and 
results in a reduced $\chi^2$ of 1.01. This single-burst model is shown in the 
bottom panel of Figure 5.

The absolute magnitudes of the three objects were determined 
using the apparent magnitudes given in Table 3, along with
$K$-corrections derived from their best-fit SED. In the $R$ band,
these are $-18.1$, $-18.6$, and $-16.6$ for objects 1, 3, and 4, respectively, giving
a total $R$-band luminosity for the galaxy of $M_R=-19.2$ or $L_R=0.16L_R^*$. 
Here, we have added the luminosities of the three features and reported 
the result as the luminosity of the DLA galaxy. In the 
$K$ band, we find absolute magnitudes $-19.7$, $-20.2$, and $-19.6$,
with a total $K$-band luminosity of $M_K^*=-21.0$ or $L_K=0.04L_K^*$.

\subsubsection{More EROs}

Closer inspection of Figure 4 reveals many objects in this $50\arcsec
\times 50\arcsec$ field that are relatively bright in $J$ but are
below the detection limits or are very faint in $R$. Several of these
are in the vicinity of galaxy b and could well be dusty regions in
which star formation is being triggered by the interaction between
galaxies a and b.  Their location is compelling evidence for
their association with the $z=0.313$ galaxies, although the possibility
that they are associated  with the radio-loud, $z_{em} = 1.187$, quasar
cannot be ruled out.  


\subsubsection{Kinematics}

High resolution observations of the \ion{Mg}{2} absorption line have not been
published,
but the high rest equivalent width of the 2796 \AA\ line, $W_0^{\lambda2796}
=2.21$ \AA, implies a velocity width of $\gtrsim 240$ km s$^{-1}$.
The 21 cm absorption line profile of this DLA system is also complex
(Lane 2000), and has been shown to vary on time scales of a few days
(Kanekar \& Chengalur 2001b). The 21 cm absorption line extends over 85 km
s$^{-1}$ and is resolved into five components. The deepest component,
with optical depth $\tau \approx 0.11$, is at the low-velocity end of
the profile. Prochaska \& Wolfe (1997, 1998) have used simulations to
show that this type of leading-edge profile is produced along a line of sight that
passes through a rotating disk. On the other hand, Haehnelt, Steinmetz, \& Rauch
(1998), who consider gas infall due to merging, and 
McDonald \& Miralda-Escud\'e (1999) who consider moving clouds in a
spherical halo, have also reproduced leading-edge line profiles. 
In the case of the PKS 1127$-$145 DLA system, both the image and kinematics 
are more consistent with the latter interpretations. 
However, this DLA system appears to be even more
complex in its structure and kinematics, since it is not an isolated
system and probably owes its morphology to the nature of its
environment.

\subsection{PKS 1629+120}

The sight line toward the quasar PKS 1629+120 ($V=18.4$, $z_{em}=1.795$) 
contains a DLA system at $z=0.532$ with column density
$N_{HI}=(5.0\pm1.0)\times10^{20}$ atoms cm$^{-2}$.  We discovered this
system in an {\it HST} Cycle 9 survey for DLA lines  in strong
\ion{Mg}{2}-\ion{Fe}{2}  absorption-line systems.
\begin{figure*}
\centerline{\epsfig{file=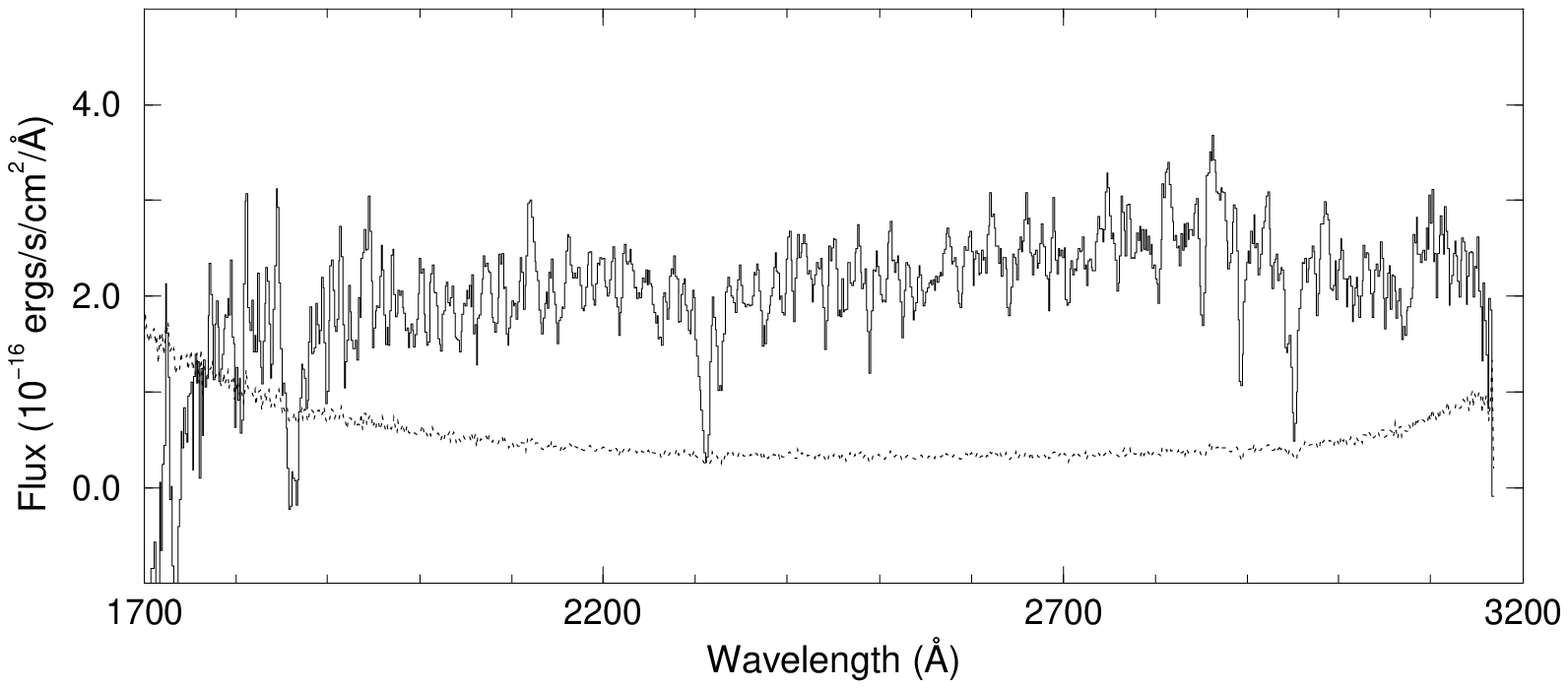,width=6.in}}
\caption{\footnotesize HST-STIS spectrum of PKS 1629+120. The $1\sigma$ error
spectrum  is also shown.  Both of the \ion{Mg}{2} lines along this line of sight
are associated with strong Ly$\alpha$ absorption. The 
absorption feature at 1863 \AA\  is a DLA line at $z=0.532$ with
$N_{HI}=(5.0\pm1.0) \times 10^{20}$ atoms cm$^{-2}$ (see Fig. 7) and
the absorption line at 2306 \AA\ is sub-DLA at $z=0.901$ with
$N_{HI}=(5.0\pm0.4)\times10^{19}$ atoms cm$^{-2}$. The
Ly$\alpha$ line at 2895 \AA\ has rest equivalent width
$W_0^{\lambda1216} =1.4$ \AA\  and is associated with a  \ion{C}{4}
system at $z=1.3786$ (Aldcroft, Bechtold, \& Elvis 1994).  The strongest line in
the blend at 2952 \AA\ has $W_0^{\lambda1216}=2.5$ \AA; it has no
known associated metal line absorption.}
\end{figure*}
\begin{figure}
\epsscale{1.0}
\plotone{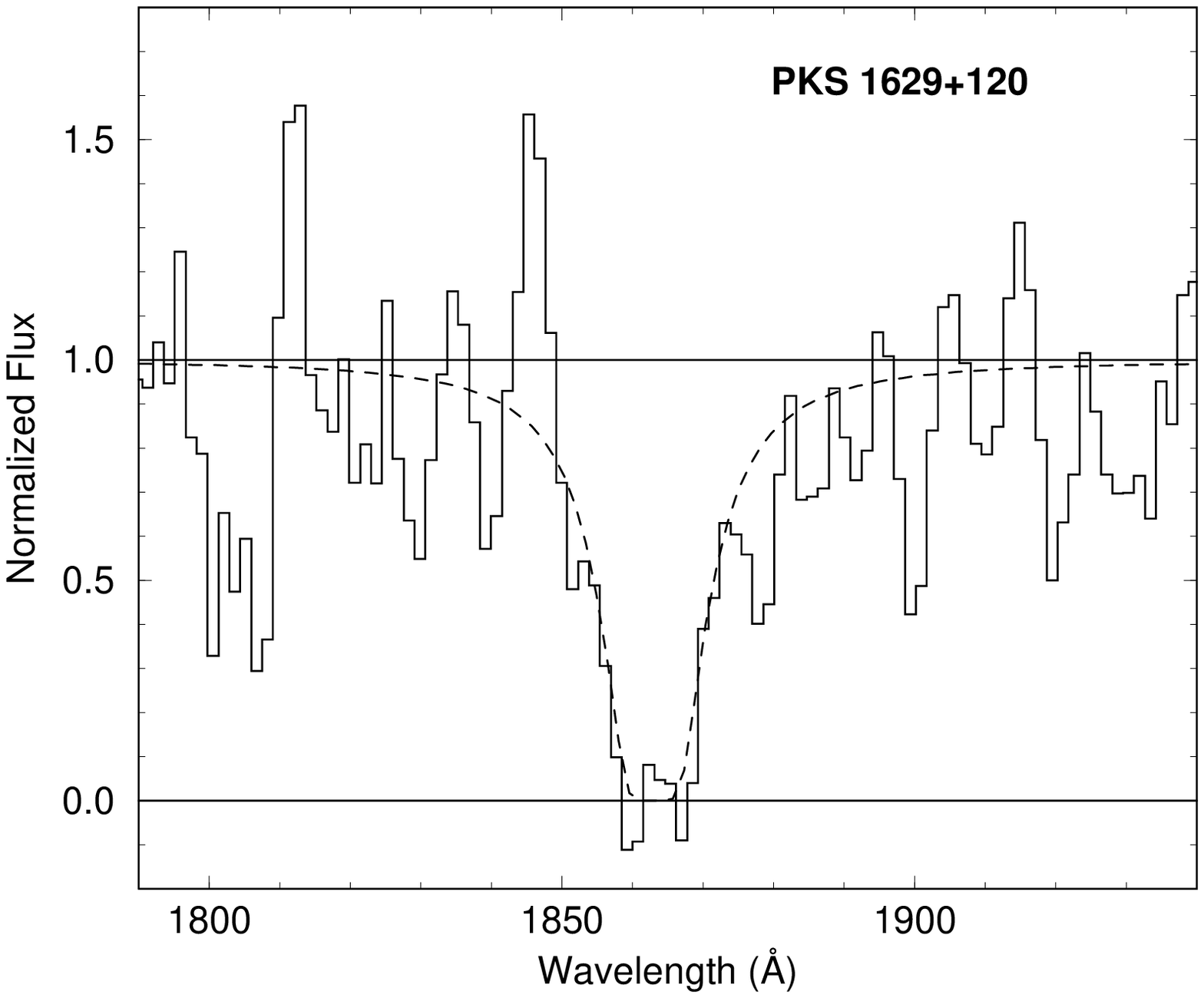}
\caption{\footnotesize A portion of the HST-STIS spectrum of PKS 1629+120 showing
the DLA line at $z=0.532$. The overlaid Voigt profile has
$N_{HI}=5.0\times10^{20}$ atoms cm$^{-2}$.}
\end{figure}

\subsubsection{The UV Spectrum}

 Figure 6 shows the
{\it HST}-STIS G230L NUV-MAMA spectrum, while  Figure 7 shows a Voigt
profile with column density $N_{HI}=5.0\times10^{20}$ atoms cm$^{-2}$
overlaid  on the DLA line. The  error in the column density estimate
is dominated by the uncertainty in continuum placement and was
determined using a procedure similar to that described in RT2000.
This line of sight has two \ion{Mg}{2} systems, one at $z=0.5313$ with
\ion{Mg}{2} $W_0^{\lambda2796}  = 1.40$ \AA\ and \ion{Fe}{2}
$W_0^{\lambda2600} = 0.70$ \AA\ (Aldcroft et al. 1994),
and one at $z=0.9005$ with \ion{Mg}{2} $W_0^{\lambda2796}  = 1.06$
\AA\ and \ion{Fe}{2} $W_0^{\lambda2600} = 0.63$ \AA\ (Barthel, Tytler,
\& Thomson 1990). The $z=0.9005$ system is sub-DLA with
$N_{HI}=(5.0\pm0.4)\times10^{19}$ atoms cm$^{-2}$.  We also detect a
Ly$\alpha$ line at 2895 \AA\ with rest equivalent width
$W_0^{\lambda1216} =1.4$ \AA\ that is associated with a  \ion{C}{4}
system at $z=1.3786$ (Aldcroft et al. 1994).  The strongest line in
the blend at 2952 \AA\ has $W_0^{\lambda1216}=2.5$ \AA; it has no
known associated metal line absorption.

\subsubsection{Imaging Results}

Figure 8 shows $\approx$40\arcsec\ $\times$ 40\arcsec\
$U$, $B$, $R$, and $K$ images of this field.  All four quasar
PSFs have been subtracted, resulting in residuals that are
comparable to the background in each image. The position of the 
quasar is marked with a plus sign. Unresolved objects are
labeled with an ``S'' and are presumed to be stars, while resolved objects 
are galaxies labeled with a ``G''. They are numbered in order of increasing
distance from the quasar. G1 appears patchy
in $U$, possibly indicating  regions of recent star formation.  An
edge-on, disklike structure can be seen in $B$. It appears bulgelike
in $K$ and shows signs of both bulge and disk structure in $R$. 
These morphological features are consistent with G1 being a mid-type
spiral galaxy. As we show below, the photometry of G1 is consistent with it
being the DLA galaxy at $z=0.532$. In this case, its impact parameter, i.e., 
the projected distance from the center of G1 to the center of the PSF of the 
quasar is $b=17$ kpc. 

\subsubsection{Photometry of G1}

Photometric measurements and  $1\sigma$ uncertainties
of G1 are given in Table 3. These magnitudes were used to determine
a photometric redshift for G1 by using the galaxy spectral templates 
of G. Bruzual \& S. Charlot (2003, in preparation) in a principal component
analysis (A. Conti 2002, private communication). The best-fit template was
determined at redshifts $0<z<2$ in steps of 0.001. The resulting 
reduced $\chi^2$, which was determined at each step, has a minimum 
equal to 0.7 at $z\approx 0.59$, and redshifts in the range $0.50<z<0.65$ are 
good fits, with reduced $\chi^2\le 1$. The reduced $\chi^2$ at 
$z=0.90$ is $\approx 4$, making $z=0.90$ approximately 30 times less probable 
than $z=0.59$. Therefore, we can rule out the possibility that 
G1 is the sub-DLA galaxy at $z=0.901$ (see \S3.4.4), and assume 
that G1 is the DLA galaxy at $z=0.532$. However, we note that a slit 
redshift is required to unambiguously determine the redshift of G1. 

We also fit stellar population synthesis models to the photometry, as
described in the Appendix. The best-fit model, which results in a
reduced  $\chi^2=0.82$, is shown in Figure 9. It is a single 0.05 Gyr
old burst  and has $E(B-V)=0.5$.  The apparent magnitudes from Table 3,
in combination with  $K$-corrections derived from this  model give
absolute magnitudes  of $M_U=-20.3$, $M_B=-20.3$, $M_R=-21.3$,
$M_J=-23.0$, and $M_K=-23.9$  at $z=0.532$.
\begin{figure*}
\centerline{\epsfig{file=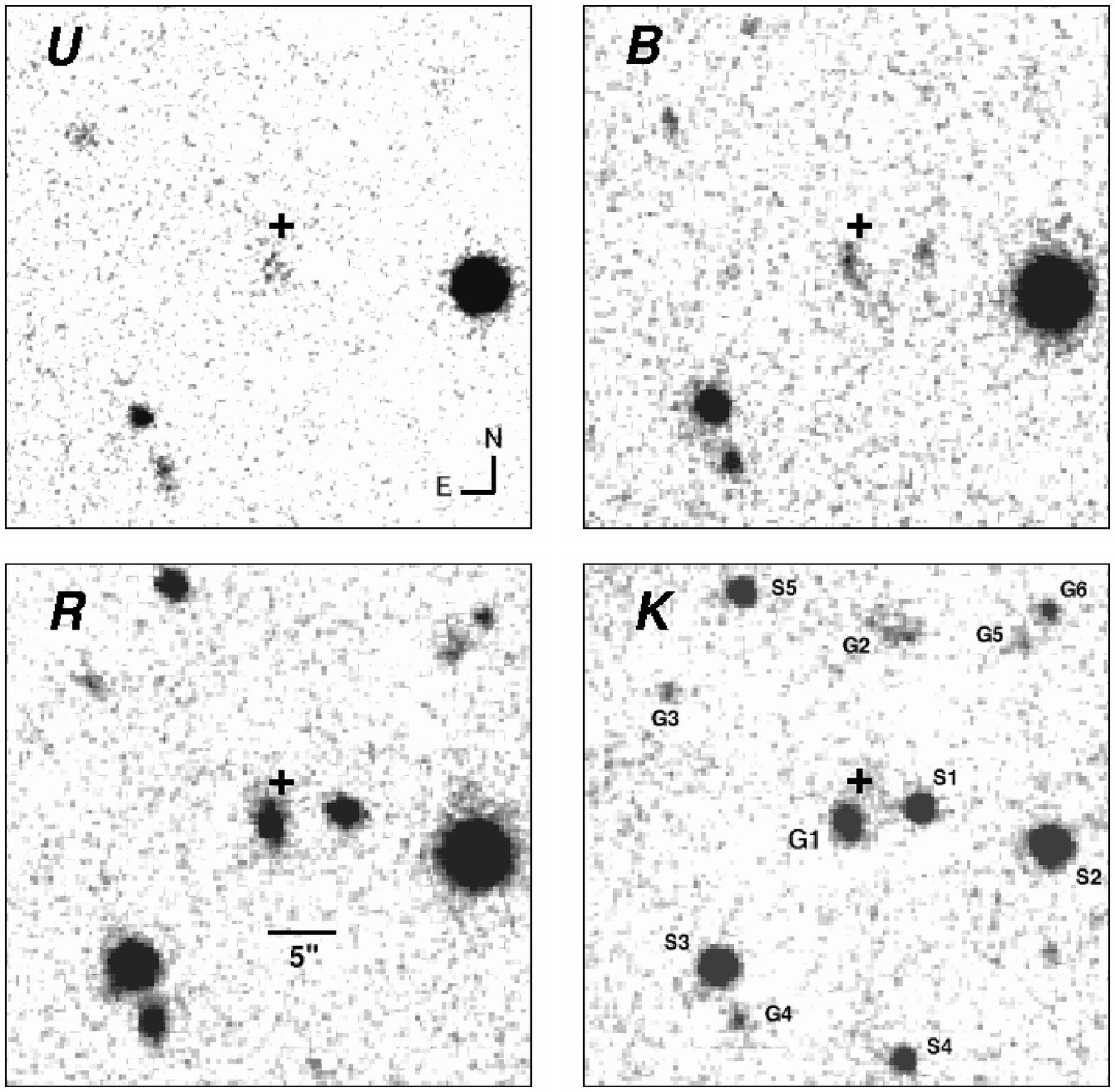,width=6.in}}
\caption{\footnotesize $U$, $B$, $R$, and $K$ images of the PKS
1629+120 field that contains a DLA system at $z=0.532$. The PSF
of the quasar, whose position is marked with a plus sign, has been
subtracted in all four images. The PSF subtraction here is cleaner
than in the other three DLA fields, since PKS 1629+120 is relatively
faint. Resolved objects are labeled with a  ``G'' and unresolved
objects are labeled with an ``S''.  G1 is identified as the DLA
galaxy. }
\end{figure*}
\begin{figure}[h]
\epsscale{1.0} \plotone{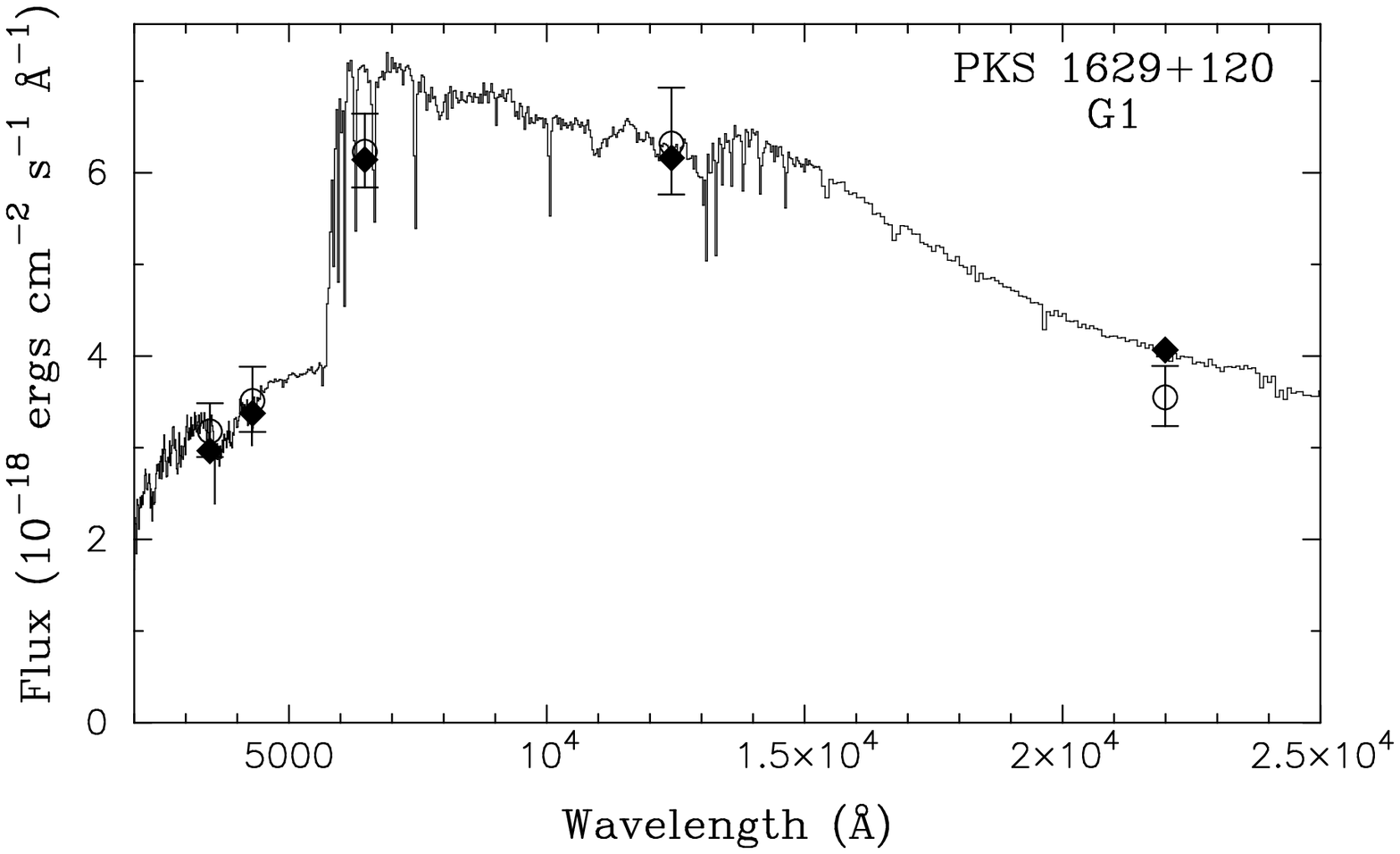}
\caption{\footnotesize Best-fit single-burst model at $z=0.532$
overlaid on the photometry for G1 in the  PKS 1629+120 field. 
Symbols are as in Fig. 2. The model is a 0.05 Gyr old single burst with
$E(B-V)=0.5$.}
\end{figure}

\subsubsection{The $z=0.901$ Sub-DLA Galaxy}

G1 is the only galaxy detected near the quasar sight line, and we have
shown that its photometry is consistent with it being the $z=0.532$
DLA galaxy. Having assumed that G1 is at
$z=0.532$, we  must now consider the whereabouts of the $z=0.901$
sub-DLA galaxy.  One possibility is that it, too, has a small impact
parameter but is below  the detection limit of all our images. A
second possibility is that it lies directly along the line  of sight
to the quasar and was subtracted along with the quasar PSF. Third,  it
might be one of the galaxies detected at large impact parameter. We
consider each of these in turn.

If the first possibility is true, then the $z=0.901$ galaxy must have a
surface brightness of $\mu_K > 21.6$ mag arcsec$^{-2}$ and 
$\mu_R > 25.4$ mag arcsec$^{-2}$ at the $3\sigma$ level (see Table 2). 
Assuming that it is a face-on galaxy with an extent of 20 kpc, we find
that $m_R > 23.2$ and $M_R > -21.5$, where a $K$-correction for an Sb-type
galaxy has been applied (Poggianti 1997), or $L_R < 1.3L_R^*$. Similarly, 
in the $K$ band we find  $m_K > 19.4$ and $M_K > -24.4$, or $L_K<0.4L_K^*$. 

The quasar PSF was subtracted reasonably cleanly, since the PSF was well
sampled by several stars brighter than the quasar in each frame. If there is 
a galaxy coincident on the sky with the quasar, then it would have to
mimic a point source and be centered exactly on the quasar PSF. Although this
is a possibility, it is highly unlikely. 

Except for G1, all detected galaxies lie between 11\arcsec\ and 19\arcsec\ from the 
quasar and have impact parameters between 72 and 124 kpc at $z=0.901$, i.e.,
much larger than what is typical for \ion{Mg}{2} absorbers (Steidel 1993). 
The implied impact parameters at $z=0.532$ are also very large, ranging between 
62 and 107 kpc. Therefore, from impact parameter
considerations, these are not likely to be either of  the two absorbing
galaxies.

Finally, we note that if G1 was at $z=0.901$, then the DLA galaxy, which would 
have gone undetected, would have to have $L_R < 0.3 L_R^*$ and $L_K < 0.2 L_K^*$. 
Moreover G1 would have $L_R=2.1L_R^*$ and $L_K=4.5L_K^*$, and it would be
an unusually luminous low-redshift sub-DLA galaxy.
Thus, it is more reasonable to conclude that G1 is at $z=0.532$
and that the $z=0.901$ sub-DLA galaxy is not detected in our images.

\section{Discussion}

The four new low-redshift DLA galaxies presented here have an
interesting mix of morphologies. The two higher redshift galaxies,
G1 B2 0827+243 ($z=0.525$) and G1 PKS 1629+120 ($z=0.532$)
are $\approx L^*$ spirals. Higher resolution {\it HST}
observations of the former reveal that an interaction between the
luminous spiral and a satellite dwarf galaxy might be responsible
for the presence of DLA gas at larger galactocentric distances.
We have, a posteriori, detected an extension to G1
in our ground-based images that might indicate the presence of 
a satellite. In the absence of high resolution images of the 
PKS 1629+120 field, we cannot offer a more detailed description 
of that DLA galaxy's morphology.  

Images of the two lower redshift DLA systems are more complex. 
Lacking higher resolution images, there are two possible identifications 
for the $z=0.239$ PKS 0952+179 DLA galaxy. Either it is the object(s) that 
overlaps with the QSO PSF in the $J$-band image,  or it is
a larger, irregular dwarf galaxy with patchy structure. In the former case, 
it is unlikely that the luminosity of this DLA galaxy exceeds a few hundredths
of $L^*$. Faint, patchy morphology is also present in
the $z=0.313$ DLA galaxy in the PKS 1127$-$145 field. However, in
this case numerous star-forming regions are clearly visible. These
regions are blue and displaced enough from the quasar sight line that
a confirming slit redshift has been obtained for one of them.
Because of the appearance of this patchy irregular structure, it
is likely that the nearby $\approx L^*$ spiral galaxies,
the closer one of which was once thought to be the DLA galaxy, are
responsible for tidally disrupting the actual DLA galaxy that lies
along the line of sight to the quasar. Thus, the DLA galaxy takes the
form of an extended irregular structure with star-forming regions.
However, the possibility that these star-forming regions are
individual interacting or merging galaxies cannot be ruled out.
Again, higher resolution images would be needed to clarify the
situation.  Whichever is the case, the total light from the DLA
galaxy does not exceed $0.16 L^{*}$.

These observations add to the small but growing list of DLA
galaxies at low redshifts.  Table 4 lists the status and properties
(luminosity relative to $L^{*}$, $L/L^{*}$, neutral hydrogen column density, 
$N_{HI}$, and impact parameter, $b$) of galaxy identifications for 14 
DLA systems with $z_{DLA} \lesssim 1$ and $N_{HI}\ge 2 \times 10^{20}$  
atoms cm$^{-2}$ (but not including local galaxies).  The distributions
of DLA galaxy properties for these 14 cases are shown in Figure
10. $B$-band luminosities are plotted when available. $K$-band
luminosities are plotted for galaxies that are not detected in $B$,
but this substitution should not grossly affect any trends in the
data.  For the case of the DLA galaxy in the 3C 336 field, only an upper
limit to its luminosity can be determined, since the galaxy is not detected.
For the same reason, the impact parameter is unknown. This case is clearly 
representative of another dwarf DLA galaxy if stars are present in the 
\ion{H}{1} gas at all. Although the data set contains only 14 systems, 
some important trends emerge.

\begin{center}
\begin{deluxetable*}{lcclclc}
\tabletypesize{\footnotesize}
\tablewidth{0pt} 
\tablecaption{Low-Redshift DLA Galaxy Properties\tablenotemark{a}} 
\tablehead{
\multicolumn{1}{c}{QSO} &  
\multicolumn{1}{c}{$z_{\rm DLA}$} &
\multicolumn{1}{c}{$N_{\rm HI}$/10$^{20}$} &
\multicolumn{1}{c}{Luminosity\tablenotemark{b}} &
\multicolumn{1}{c}{$b$} &  
\multicolumn{1}{l}{Morphology} &
\multicolumn{1}{c}{Reference}\\[.2ex] 
\colhead{} & 
\colhead{}  &
\multicolumn{1}{c}{atoms cm$^{-2}$} & 
\colhead{}  &
\multicolumn{1}{c}{kpc} & 
\colhead{} & 
\colhead{}  } 
\startdata
\multicolumn{7}{c}{This work}\\ [0.2ex]
\hline\\
B2 0827$+$243        &  0.525 & 2.0 & $0.8L_B^*$, $1.6L_R^*$, $1.2L_K^*$ & 34 & Disturbed spiral & 1 \\ 
PKS 0952$+$179       &  0.239 & 21  & $0.02L_K^*$\tablenotemark{c} & $<4.5$  & Dwarf LSB  & 1 \\ 
PKS 1127$-$145       &  0.313 & 51  & $0.12L_B^*$, $0.16L_R^*$, $0.04L_K^*$\tablenotemark{d} & $<6.5$ & Patchy/irr/LSB & 1 \\ 
PKS 1629+120         &  0.532 & 5.0 & $0.6L_B^*$, $1.1L_R^*$, $0.6L_K^*$ & 17 & Spiral & 1 \\[0.2ex]
\hline\\
\multicolumn{7}{c}{Other work}\\ [0.2ex]
\hline\\
AO 0235+164          &  0.524 & 45 & $0.8L_B^*$   & 6.0   & Late-type spiral\tablenotemark{e}&2,3\\
EX 0302$-$223        &  1.010 & 2.3 & $0.2L_B^*$   & 9.2   & Semi-compact & 4 \\ 
PKS 0454$+$039       &  0.859 & 4.7 & $0.4L_B^*$   & 6.4   & Compact & 4 \\ 
Q0738$+$313 (OI 363) &  0.091 & 15  & $0.08L_K^*$  & $<3.6$& LSB &  5 \\ 
\nodata              &  0.221 & 7.9 & $0.1L_B^*$   & 20    & Dwarf spiral & 5 \\
Q0809+483 (3C 196)   &  0.437 & 6.3 & $1.7L_B^*$   & 9.6   & Giant Sbc & 4 \\
Q1209$+$107          &  0.633 & 2.0 & $1.6L_B^*$   & 11.2  & Spiral & 4  \\ 
PKS 1229$-$021       &  0.395 & 5.6 & $0.1L_B^*$   & 7.6   & LSB & 4,6  \\ 
Q1328+307 (3C 286)   &  0.692 & 15  & $0.4L_B^*$   & 6.5   & LSB & 4,6   \\ 
Q1622$+$239 (3C 336) &  0.656 & 2.3 & $<0.05L_K^*$ &\nodata& LSB? compact? & 7 \\ 
\enddata
\tablenotetext{a}{Only galaxies with cosmological redshifts are included. Thus,
the DLA galaxy at $z=0.010$, SBS 1543+593, is excluded since it does not fall well
beyond the local velocity field, which is defined at 3000 km/s by the outer boundary 
of the 
Virgo cluster (Binggeli, Popescu, \& Tammann 1993). We note that it is an LSB galaxy with
 $L_B=0.02L_B^*$ and $b=0.8$ kpc (Bowen, Tripp, \& Jenkins 2001; 
Schulte-Ladbeck et al. 2002).}
\tablenotetext{b}{$M_B^*=-20.9$ (Marinoni et al. 1999), $M_R^* = -21.2$ (Lin et al. 1996), 
and $M_K^*=-24.5$ (Loveday 2000), for $q_0=0.5$ and $H_0=65$ km s$^{-1}$ Mpc$^{-1}$.}
\tablenotetext{c}{Sum of luminosities of objects 1 and 2  (see \S3.2.2 and Figure 3).}
\tablenotetext{d}{Sum of luminosities of objects 1, 3, and 4 (see \S3.3.1 Figure 4).}
\tablenotetext{e}{The object defined as A1 in Yanny, York, \& Gallagher 
1989 and Burbidge et al. 1996
has the smallest impact parameter and, therefore, we have assumed that it is the 
DLA galaxy.
A $K$-correction of 1 magnitude, which is appropriate for a late-type spiral galaxy
at $z=0.524$ (Poggianti 1997) has been applied to derive the absolute luminosity of A1
from the apparent magnitude measured by Burbidge et al. 1996. The HI column
density is from Turnshek et al. 2003.}
\tablerefs{(1) This paper. 
(2) Burbidge et al. 1996.
(3) Yanny et al. 1989.
(4) Le Brun et al. 1997. 
(5) Turnshek et al. 2001.  
(6) Steidel et al. 1994. 
(7) Steidel et al. 1997.}  
\end{deluxetable*}
\end{center}

\begin{figure*}
\centerline{\epsfig{file=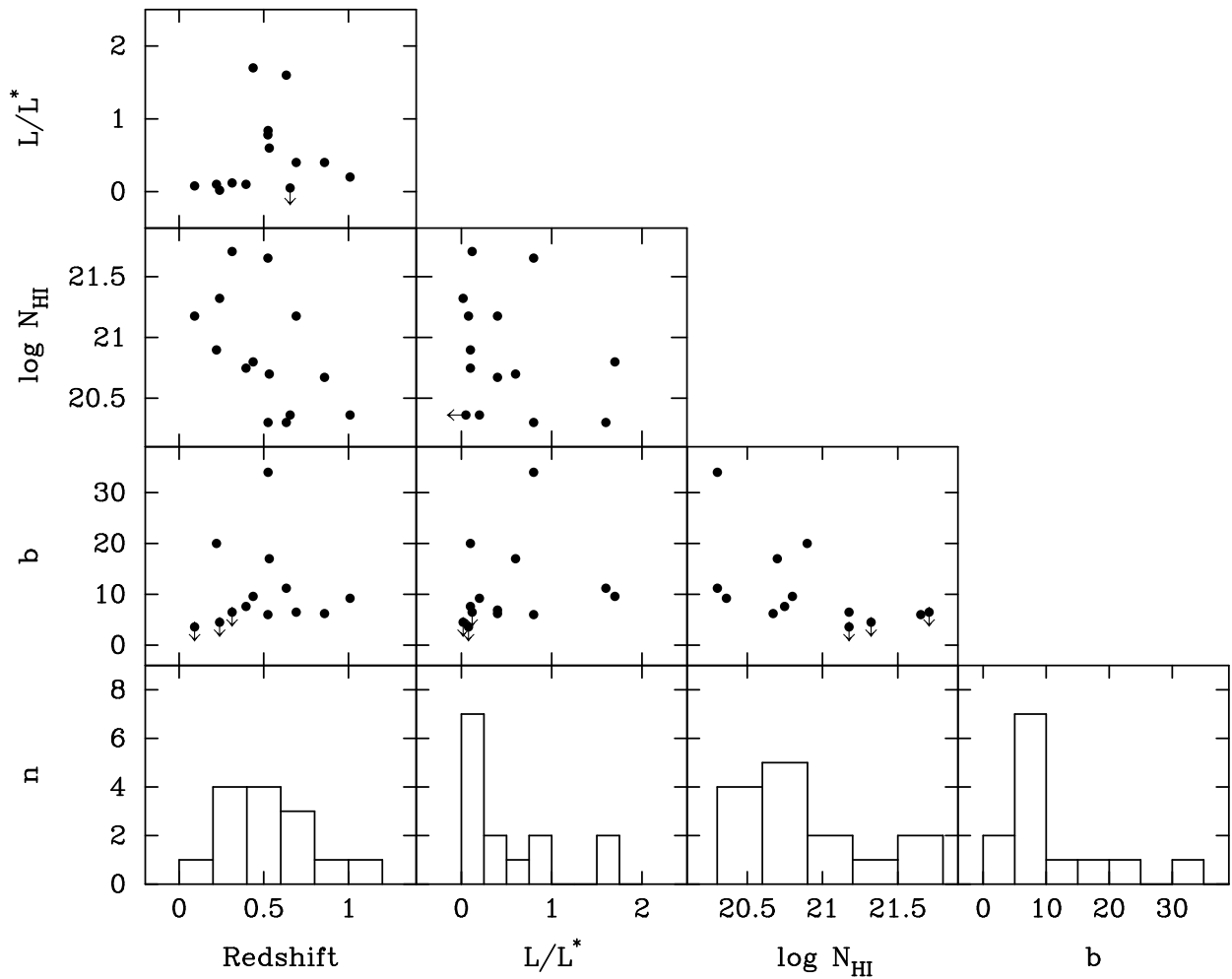,width=6.5in}}
\caption{\footnotesize Distribution of properties for the DLA galaxies listed 
in Table 4. An upper limit for the luminosity of the DLA galaxy in the 
3C 336 field is plotted, but an impact parameter is not, since no DLA galaxy has 
been detected. $B$-band luminosities are plotted in all cases except the two for 
which only $K$-band measurements exist. The impact parameter $b$ is in kiloparsecs.
Upper limits in $b$ are indicated by arrows. The $b$ histogram has
upper limits included as measurements.}
\end{figure*}
The $b$ vs. $L/L^*$ plot shows that low-luminosity dwarf galaxies
with small impact parameters dominate this small sample.  Since DLA
galaxies are \ion{H}{1} cross section--selected, this means that
sub-$L^*$ dwarf galaxies dominate the \ion{H}{1} cross section at
$z\approx0.5$. This is also indicated by the luminosity histogram,
$n$ versus $L/L^*$, which implies that it is $\approx 3$ times more
likely for a quasar sight line to intercept an $\approx 0.1L^*$
galaxy than an $\approx L^*$ galaxy.  For galaxies at the present epoch,
 Zwaan, Briggs, \& Verheijen (2002, see their Fig. 1) show that 
the probability of intercepting DLA gas, i.e., the cross-sectional
area of DLA gas per unit volume, as a function of galaxy magnitude is 
flat for a steeply rising optical luminosity function with $\alpha =
-1.5$ and decreases at faint magnitudes for $\alpha=
-1.2$ or $-1.0$. Expressed as a function of luminosity, this
probability would rise as $\sim L^{-1}$ toward the faint end for the 
case in which $\alpha = -1.5$ and, based on the data points in their Figure 1,
would increase as $\sim L^{-0.5}$ toward the faint end 
for  $\alpha=-1.0$. For the latter case, the trends at $z=0$ and 
$z\approx 0.5$ are similar. Turnshek, Rao, \& Nestor
(2002) arrived at much the same conclusion (see their Fig. 4). 
They compared the optical diameter-limited ($>$7\arcmin)
complete sample of $z=0$ galaxies from Rao \& Briggs (1993) 
to a $z\approx 0.5$ sample similar to the one in Table 4. They found that
the relative absolute magnitude distributions of the two samples are
comparable within the large uncertainties associated with
 the low-redshift sample.\footnote{This statement is based on a relative
comparison at the two epochs, independent of any 
evolution in $dn/dz$ (which measures evolution in galaxy number density 
times galaxy cross section) for DLA systems (RT2000).} We note that
Table 4 has our final measurements for the four DLA galaxies discussed in
this paper, as opposed to the preliminary measurements used in Turnshek
et al. (2002). Also, the galaxy 
toward AO 0235+164 was not included in their analysis. 

The trend that higher column densities tend to be observed at low
impact parameters ($b$ vs. $\log N_{HI}$) is not unexpected. However,
what is unexpected is that the highest column densities are
observed in galaxies with the lowest luminosities ($\log N_{HI}$ vs. 
$L/L^*$). This latter trend might be explained if a selection bias
exists such that bright galaxies with high column densities at
small impact parameters are missing from our sample. This might
be caused by dimming of the background quasar due to reddening
from the foreground DLA gas, excluding such bright quasars from
our magnitude-limited sample. While a study by Ellison et al. (2001) on the
occurrence of DLA systems at $z\approx 2$ in a radio-selected quasar sample did not show
a significant excess in the DLA number density as compared to DLA systems
found in optically selected quasar samples, this might not hold true
at lower redshift. 

Two specific oddities of the sample 
are worth mentioning here: (1) Four of the five  DLA galaxies with
$N_{HI}>1\times 10^{21}$ atoms cm$^{-2}$ are LSB galaxies. This
may have significant implications for the whereabouts of the bulk of 
the \ion{H}{1} gas, since the determination of $\Omega_{HI}$
from DLA systems is dominated by ones with the highest column densities. 
These four DLA systems contain 55\% of the total column density in 
Table 4. Zwaan et al. (2003) have shown that
only 15\% of the neutral gas at the present epoch is contained in 
LSB galaxies. Thus, if the low-redshift DLA galaxy
trends hold up with larger samples, it would indicate that a different
population of objects is responsible for the
bulk of the neutral hydrogen gas in the universe at $z \approx 0.5$.
(2) The three galaxies that have the largest impact parameters (17,
20, and 34 kpc) are spirals. G1 in the B2 0827+243 field is
an $\approx L^*$ spiral and has the largest impact parameter, and, as
discussed earlier, it shows signs of an interaction that might
be responsible for distributing the gas to larger galactocentric
distances.  It was only possible to see evidence for this in
the high-resolution {\it HST} image that exists for the field.
Similar high-resolution {\it HST} images do not exist for the
other two large-impact parameter fields, i.e., the dwarf spiral galaxy in
the OI 363 field and the $\approx L^*$ spiral galaxy in the PKS 1629+120 field.
Study of these two fields at higher resolution would be highly
desirable in order to understand the reality and nature of this
DLA gas at unusually large galactocentric distance.

\acknowledgements

We are grateful to the staff at the NOAO, and to the WIYN queue team 
in particular, for their assistance with the observing. We also thank
the MDM observatory and IRTF support astronomers for their assistance. 
The ground-based component of this work has been funded in part
by an NSF grant, while the space-based component of this work
has been funded in part by a NASA LTSA grant. Support was also
provided by NASA through a grant from the Space Telescope Science
Institute, which is operated by AURA, Inc., under NASA contract NAS 5-26555.
Alberto Conti is gratefully acknowledged for using his galaxy template
PCA code to derive a photometric redshift for G1 in the PKS 1629+120 field.
D. B. N. acknowledges partial support from a Daniels graduate student
fellowship and a Mellon graduate student fellowship at the University
of Pittsburgh. 

\appendix

\section{Stellar Population Template Fits}

In order to investigate the stellar populations present in
the  DLA galaxies, we fitted stellar population
synthesis models to the photometry when sufficient multi-band
data were available. However, multiple stellar populations
typically contribute to an observed 
SED.  Therefore, we used a fitting method that allows for
contributions from multiple stellar populations attenuated by a
variable amount of wavelength-dependent extinction due to dust.
At the same time, galaxy colors are known to be degenerate in
age-metallicity-redshift space, so further assumptions and an
interpretation of what our fits mean are generally required.

First, when performing fits, we set the redshift of the
galaxy to the DLA redshift.  In all three cases there was reasonable
justification for this.  The adopted redshift was either
confirmed spectroscopically or verified to be a likely photometric
redshift.  Second,
the metallicities of all the template spectra were fixed to be
solar. On the one hand, this is useful, since solar metallicity is
an often-used benchmark. However, it is well known that the measured
metallicities of DLA gas at moderate-to-high redshift are closer
to 1/10 solar. If lower metallicities (e.g.,  $\approx 0.2 -
0.02$ solar) hold  for the DLA galaxies studied here, the nature
of the age-metallicity degeneracy means that we would generally
underestimate the age of the stellar population(s) in the DLA
galaxy. If the stellar metallicity were $\approx 0.2$ solar,
the effect would be quite small, but at metallicities as low as
$\approx 0.02$ solar, the effect is substantial.  Thus, even with
the advantage of fixing redshift, the age-metallicity degeneracy
gives rise to some interpretive limitations.  Another limitation
is the degree to which any adopted set of template spectra (see
below) are applicable to fit the observed SEDs. In any case, when
the qualitative results of the SED fitting are taken in combination
with morphological information, we are generally able to draw a
consistent conclusion that provides useful information on the
evolutionary history of a DLA galaxy as presented in \S3.

For the template spectra we use the galaxy isochrone synthesis
spectral evolution library (GISSEL99) of G. Bruzual \& S. Charlot
(2003, in preparation). We chose templates with a Scalo initial mass function
(IMF) corresponding to
instantaneous bursts observed at eleven different ages: $10^{-3}$,
$10^{-2}$, 0.05, 0.1, 0.2, 0.3, 0.6, 1.0, 1.5, 4, and 12 Gyr.
From these original 11, we generated an additional 110 templates
by applying a Calzetti reddening law (Calzetti et al. 2000) with 10
different dust extinction values, corresponding to $E(B-V)$ of 0.05,
0.1, 0.2, 0.3, 0.4, 0.5, 0.6, 0.7, 0.8, and 0.9.  The convolution
of each of these 121 redshifted templates with the UBRJK filter set
(UBRIK for B2 0827+243) was converted to flux units, resulting in
a flux 5-vector for each template.  Models were then built to
minimize the $\chi^2$ fit to the observed photometric SEDs for
each of the $121!/(121-N)!N!$ combinations of $N$ 5-vectors,
where $N$ was the number of component templates used in the fit. We
investigated cases in which the value of $N$ was 1, 2, or 3. The model
that gave the most reasonable value for reduced $\chi^2$ was then
reported as the most likely (combination of) stellar population(s)
in a galaxy.  For example, both single- and two-burst models are
reported as best fits to the photometry of G1 in the B2 0827+243
field, while a single-burst model  was clearly appropriate
for fitting the photometry of G1 in the PKS 1629+120 field
(i.e., a two-burst model had a noticeably larger reduced $\chi^2$).
Although we explored three-burst models, the resulting improvements
in reduced $\chi^2$ values did not justify their use.

The standard interpretation of reduced $\chi^2$ fitting results holds
for the best fits presented in \S3.  That is, if the adopted template
spectra are representative of our observed SEDs and if our derived
observational errors are valid, model fits that have a reduced
$\chi^2$ between $\approx$ 0.8 and 2 (which correspond to reduced
$\chi^2$  probabilities between $\approx$ 0.5  and 0.2) are taken to
be acceptable fits.  On the other hand, a reduced $\chi^2$ value much
larger than this should formally be rejected. There are two possible
reasons for this.  Either (1) the errors have been underestimated,
which might be the case if some unknown systematic error in some of
the photometry exists, or (2) the set of template spectra used to
fit the observed SEDs is inappropriate. We suspect that in cases
in which we have a larger reduced $\chi^2$, say $\approx$ 5, the quoted
fits still provide qualitative information on the nature of the DLA
galaxy's stellar population. It might be that the $\chi^2$ could be
reduced further by assuming some specific form for time-variable
star formation, and/or by using different metallicities, and/or
by tracking down an unknown systematic error in the photometry.
Another possibility is finding a reduced $\chi^2$ value much less
than one (resulting in a probability greater than 0.5).  This may
be indicative of overfitting the data or error estimates that are
too large.

The specifics of the various fits to DLA galaxy SEDs are discussed
in \S3. Models that included a negative population were rejected as
unphysical.  Quoted fractions are by burst mass, although one should
keep in mind that any such method is sensitive only to the light
output of the galaxy, and as younger populations will have smaller
mass-to-light ratios, large old populations may not be resolved
by this technique when a large young burst is present.  As noted
above, instantaneous bursts may be unreliable approximations for star
formation histories of real galaxies in many cases.  By building the
best-fit model from multiple templates, however, we in effect mimic a
nonparametric star formation history.  Nonetheless, we repeated the
process with sets of exponentially decreasing star formation rates.
The results from the two methods were consistent.  For example,
a galaxy that showed an equal combination of old and young stellar
populations from our first method would show a long {\it e}-folding time
in the second method, while a galaxy with a predominately young
population would show a very short {\it e}-folding time.


\end{document}